\documentclass[pdflatex,sn-mathphys-num]{sn-jnl}% Math and Physical Sciences Numbered Reference Style
\usepackage{graphicx}%
\usepackage{multirow}%
\usepackage{amsmath,amssymb,amsfonts}%
\usepackage{amsthm}%
\usepackage{mathrsfs}%
\usepackage[title]{appendix}%
\usepackage{xcolor}%
\usepackage{textcomp}%
\usepackage{manyfoot}%
\usepackage{booktabs}%
\usepackage{algorithm}%
\usepackage{algorithmicx}%
\usepackage{algpseudocode}%
\usepackage{listings}%
\usepackage{url}%
\definecolor{RthreeDarkGreen}{RGB}{0,100,0}

\newcommand{\revSJTU}[1]{#1}

\theoremstyle{thmstyleone}%
\theoremstyle{thmstyletwo}%

\theoremstyle{thmstylethree}%


\begin{document}
\hypersetup{colorlinks=true,allcolors=black}

\title[Observing Earth's ionospheric irregularities from the Moon]{Observing Earth's ionospheric irregularities from the Moon}

\author*[1]{\fnm{Rong} \sur{Yang}}\email{rongyang@sjtu.edu.cn}
\author[1]{\fnm{Zhihong} \sur{Li}}\email{2985606018@sjtu.edu.cn}
\author[2]{\fnm{Andrea} \sur{Nardin}}\email{andrea.nardin@polito.it}
\author[2]{\fnm{Alex} \sur{Minetto}}\email{alex.minetto@polito.it}
\author[2]{\fnm{Fabio} \sur{Dovis}}\email{fabio.dovis@polito.it}
\author[3]{\fnm{Claudio} \sur{Cesaroni}}\email{claudio.cesaroni@ingv.it} 
\author[3]{\fnm{Luca} \sur{Spogli}}\email{luca.spogli@ingv.it}
\author[3]{\fnm{Lucilla} \sur{Alfonsi}}\email{lucilla.alfonsi@ingv.it}
\author[1]{\fnm{Xirui} \sur{Miao}}\email{miaoxirui@sjtu.edu.cn}
\author[1]{\fnm{Yuquan} \sur{Ma}}\email{mayuquan@sjtu.edu.cn}
\author[1]{\fnm{Wei} \sur{Gao}}\email{gaowei1515@sjtu.edu.cn}
\author[4]{\fnm{Li-Ta} \sur{Hsu}}\email{lt.hsu@polyu.edu.hk}
\author[1]{\fnm{Xingqun} \sur{Zhan}}\email{xqzhan@sjtu.edu.cn}

\affil [1]{\orgdiv{School of Aeronautics and Astronautics}, \orgname{Shanghai Jiao Tong University}, \orgaddress{\street{800 Dongchuan Road}, \city{Shanghai}, \postcode{200240}, \country{China}}}

\affil [2]{\orgdiv{Department of Electronics and Telecommunications}, \orgname{Politecnico di Torino}, \orgaddress{\street{Corso Duca degli Abruzzi}, \city{Torino}, \postcode{10129}, \country{Italy}}}

\affil[3]{\orgdiv{Aerospace Engineering Sciences Department}, \orgname{Istituto Nazionale di Geofisica e Vulcanologia}, \orgaddress{\street{Via di Vigna Murata 605}, \city{Roma}, \postcode{00143}, \country{Italy}}} 

\affil [4]{\orgdiv{Department of Aeronautical and Aviation Engineering}, \orgname{The Hong Kong Polytechnic University}, \orgaddress{\street{Hung Hom}, \city{Hong Kong}, \postcode{999077}, \country{China}}}

\abstract{{Radio signals that propagate through the Earth's ionosphere carry information about possible irregular plasma structures. Such irregularities manifest themselves as amplitude signatures that depend on the propagation and observation geometry \cite{Morton2020IonosphericEffects,Rino1979WeakScatter}.} Conventional ground and low-Earth-orbit (LEO) measurements sample these effects close to Earth. {Here we present a systematic characterization of terrestrial ionospheric irregularities using Global Navigation Satellite System (GNSS) amplitude observations from the lunar surface, \revSJTU{at distances of approximately 356,000--398,000~km from Earth's surface} \cite{Parker2026Navigation}.} Using nearly 93~h of 1-Hz lunar-surface observations of GPS L1/L5 and Galileo E1/E5a collected by the Lunar GNSS Receiver Experiment (LuGRE), we reconstruct the propagation and scanning geometry to characterize {the spatiotemporal signatures of amplitude and phase fluctuations}. At lunar distance, the Fresnel scale approaches its finite-source limit, extending the observations to a propagation regime beyond conventional near-Earth configurations. Detrended receiver-reported carrier-to-noise density ratios ($\delta C/N_0$) reveal pronounced polar amplitude fluctuations at both frequency bands, also visible in the original measurements. Within the regions and local times sampled, large fluctuations occur in about {30\% of valid one-minute polar windows}, compared with less than 5\% in {either the equatorial or mid-latitude band}. These findings  {support using} the lunar surface as a vantage point for {studying Earth's polar ionospheric irregularities} and testing how propagation and scanning relative to the magnetic field shape their amplitude signatures. Future lunar observations could extend this approach to long-term monitoring of {Earth's ionospheric irregularities}.}

\maketitle

\section{Introduction}\label{sec1}

% R3_INTRODUCTION_P01
{
The lunar surface provides a distant vantage point for studying Earth's ionospheric irregularities through the radio signals they perturb. \revSJTU{Spatial variations in ionospheric electron density introduce phase perturbations across a propagating radio wavefront, which can subsequently give rise to amplitude fluctuations through diffraction and refractive focusing or defocusing \cite{Morton2020IonosphericEffects}.} These fluctuations depend on the plasma structure and on the propagation path. Signals received on the Moon have travelled approximately $3.8\times10^5$~km beyond Earth, extending ionospheric amplitude observations far beyond conventional near-Earth geometries.
}

\begin{figure}[tbp]
\centering
\includegraphics[width=\textwidth]{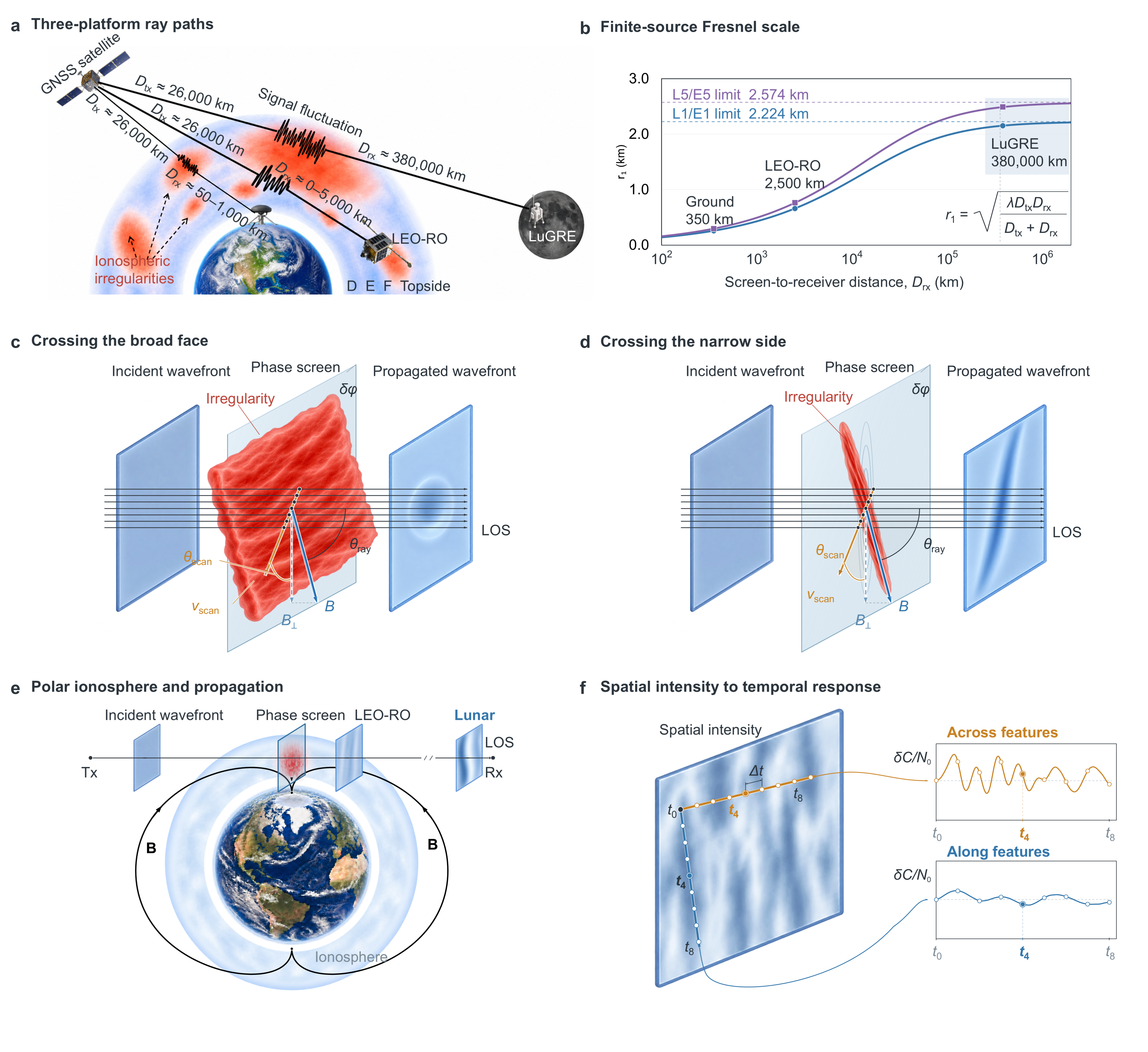}
\small
\caption{{
\textbf{Propagation and sampling geometry of lunar GNSS observations.}
\textbf{a}, The three platforms have similar transmitter-to-ionosphere distances ($D_{\rm tx}$) but vastly different ionosphere-to-receiver distances ($D_{\rm rx}$). The illustrated ground path crosses the ionosphere vertically, whereas LEO-RO and lunar paths provide limb scans, with lunar reception extending coverage to higher tangent altitudes.
\textbf{b}, For L1/E1 and L5/E5a, the first Fresnel-zone radius $r_1$ increases from \revSJTU{several hundred meters scale} for the illustrated ground and LEO-RO links to over 2 kilometres at lunar distance, approaching the finite-source limit.
\textbf{c,d}, Broad-face and narrow-side crossings through the same anisotropic irregularity illustrate how crossing geometry shapes the path-integrated GNSS signal phase perturbation $\delta\phi$. $\theta_{\rm ray}$ is the ray--magnetic-field angle; $\theta_{\rm scan}$ is the angle between transverse scan velocity $v_{\rm scan}$ and $\mathbf B_\perp$, the magnetic-field projection onto the plane normal to the line of sight (LOS).
\textbf{e}, \revSJTU{Diffraction and interference during propagation convert phase perturbations imposed by field-aligned irregularities into spatial intensity fluctuations, illustrated at LEO-RO and lunar distances for a northern polar limb path nearly perpendicular to the magnetic field $\mathbf B$.}
\textbf{f}, Relative scan direction and speed map the spatial intensity pattern into different temporal responses, illustrated for scans across or along the features as carrier-to-noise density ratio fluctuations ($\delta C/N_0$).
}}
\label{fig1}
\end{figure}

% R3_INTRODUCTION_P02
{
Ground-based Global Navigation Satellite System (GNSS) networks measure path-integrated delays and scintillation from below the ionosphere \cite{Mrak2023GroundInfrastructure}. Low-Earth-orbit radio occultation (LEO-RO) provides electron-density profiles and high-rate amplitude and phase observations along limb paths \cite{Kursinski1997GPSRO,Anthes2011ROReview,Chang2025PlanetiQSpireRO}. Coherent and incoherent scatter radars provide independent information on irregularities and background plasma within their fields of view \cite{Nishitani2019MidlatitudeSuperDARN,McCrea2015EISCAT3D}. Lunar reception adds long limb-grazing paths observed from a receiver far outside the ionosphere (Fig.~\ref{fig1}a).
}

% R3_INTRODUCTION_P03
{
The long path beyond the ionosphere changes the spatial scales that contribute to amplitude fluctuations. For a transmitter at finite distance, the first Fresnel-zone radius at an ionospheric phase screen grows towards a limiting value as the receiver recedes \cite{Rino1979WeakScatter,Rino2011GPSScintillation}. At lunar distance, it approaches the limit set by wavelength and transmitter-to-screen distance, reaching a few kilometres in the illustrated GNSS geometry (Fig.~\ref{fig1}b). Some kilometre-scale phase structures can therefore produce appreciable amplitude signatures at lunar distance while yielding weak variations along shorter propagation paths (Fig.~\ref{fig1}e).
}

% R3_INTRODUCTION_P04
{
Orientation also matters because ionospheric irregularities are often elongated along the magnetic field. Different ray directions through the same anisotropic structure produce different accumulated phase patterns and transverse gradients (Fig.~\ref{fig1}c,d) \cite{Morton2020IonosphericEffects,Rino1979WeakScatter}. The direction and speed with which the ray samples the propagated intensity pattern \revSJTU{influence how spatial structure contributes to temporal variation} (Fig.~\ref{fig1}f). These effects motivate examining how propagation and scanning relative to the magnetic field shape the lunar amplitude response.
}

% R3_INTRODUCTION_P05
{
The Lunar GNSS Receiver Experiment (LuGRE), aboard Firefly's Blue Ghost-1 lander, recorded nearly 93~h of 1-Hz GPS L1/L5 and Galileo E1/E5a observations during lunar-surface operations in March 2025 \cite{Parker2022LuGRE,Parker2026Navigation}. Recent studies used these measurements to estimate ionospheric and plasmaspheric electron content \cite{Cesaroni2026ESSOAr}, identify dispersive occultations \cite{Minetto2026IONOccultation} and model plasma delays \cite{IiyamaGao2025LunarGNSSDelays}. The amplitude signatures of terrestrial ionospheric irregularities and their relationship with lunar observing geometry have yet to be systematically characterized.
}

% R3_INTRODUCTION_P06
Here we present the first systematic characterization of terrestrial ionospheric irregularities using GNSS amplitude observations from the lunar surface. Detrended carrier-to-noise density ratios ($\delta C/N_0$) reveal stronger and more frequent polar fluctuations than at equatorial and mid-latitudes within the regions and local times sampled. The enhancement is evident in the original dual-frequency measurements and in fluctuation power above 0.1~Hz. Reconstructing the propagation and scanning geometry places these fluctuations in their observing context. These measurements extend the study of terrestrial ionospheric irregularities to lunar distances, where the Fresnel scale approaches its finite-source limit, and establish a basis for sustained observations from the Moon.

\section{Results}\label{sec2}
\begin{figure}[tbp]
\centering
\includegraphics[width=\textwidth]{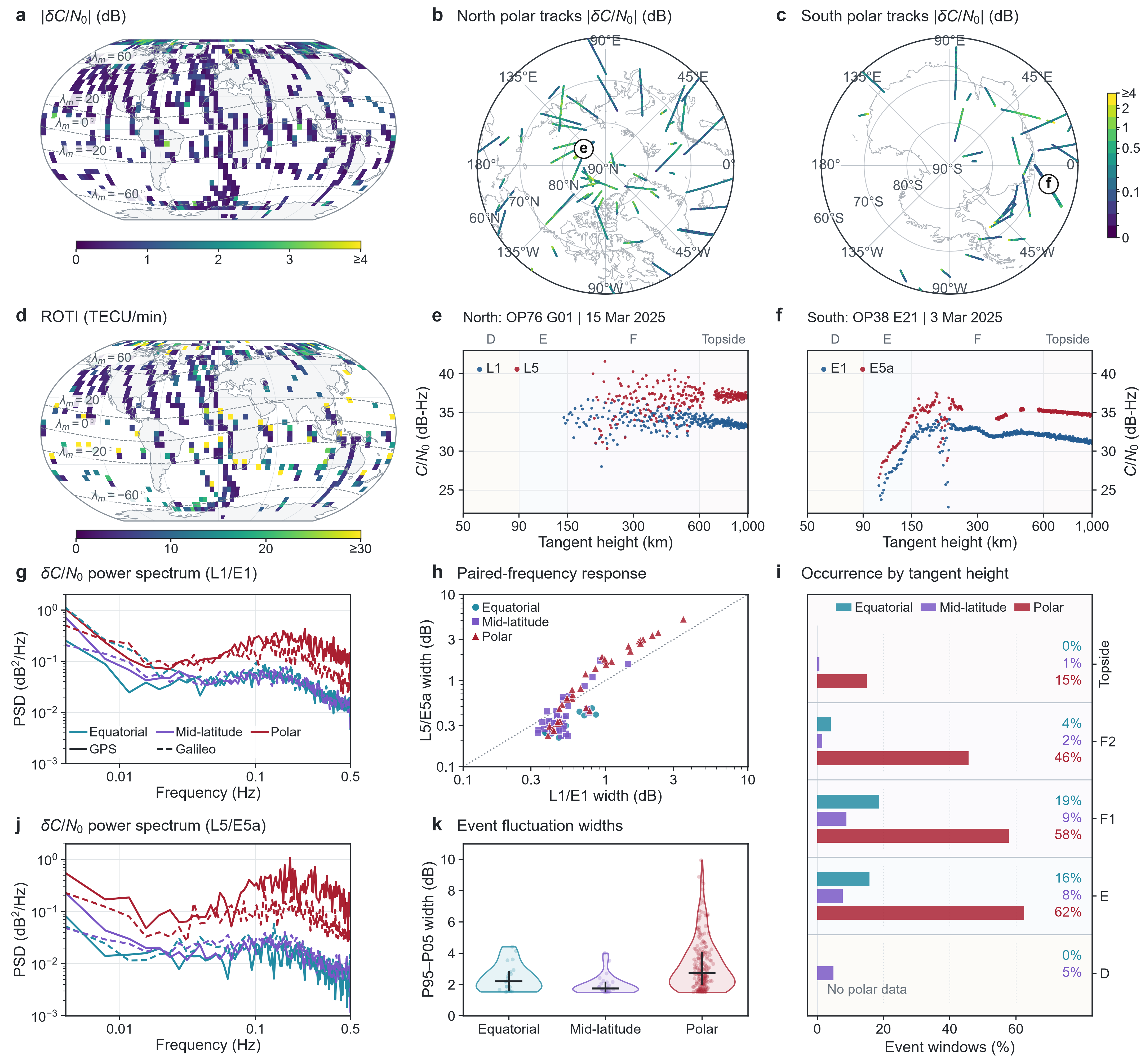}
\small
\caption{\textbf{Lunar-surface GNSS observations reveal stronger and more frequent polar amplitude fluctuations.}
 Large amplitude fluctuations cluster along tracks in both polar hemispheres, whereas enhanced ROTI spans a broader latitude range. Global maps show cell-wise P95 of 1-Hz GPS L1/Galileo E1 $|\delta C/N_0|$ (\textbf{a}) and 1-min ROTI derived from dual-frequency carrier phase (\textbf{d}) on $5^\circ\times5^\circ$ tangent-point grids; the polar views resolve individual GPS L1/Galileo E1 amplitude fluctuations along northern and southern tangent tracks (\textbf{b,c}).
 Northern GPS G01 L1/L5 records show pronounced fluctuations at F-region to lower-topside tangent heights, approximately 150--650~km (\textbf{e}). Southern Galileo E21 E1/E5a records show strong fluctuations and deep fades concentrated at F1-region tangent heights, approximately 150--250~km (\textbf{f}).
Polar median fluctuation power is greater above 0.1~Hz for GPS L1/Galileo E1 (\textbf{g}) and GPS L5/Galileo E5a (\textbf{j}).
The strongest polar paired responses generally have broader L5/E5a than L1/E1 fluctuations.  
Polar event fractions are higher across the E-region to topside tangent-height categories (\textbf{i}), and polar event widths reach approximately 10~dB (\textbf{k}). Width $W$ denotes $P_{95}-P_{05}$ of $\delta C/N_0$; event windows have a width $\geq1.5$~dB within each 1~min. Bars in (\textbf{k}) summarize medians and interquartile ranges.}
\label{fig2}
\end{figure}

%%%%%%%%%%%%%%%%%%%%%%%%%%%%%%%%%%Fig.2%%%%%%%%%%%%%%%%%%%%%%%%%%%%%%%%%%%%%%%%%%%%%%%%%%%%
% R3_RESULTS_FIG2_P01
{
Pronounced amplitude fluctuations occurred along lunar GNSS paths through both polar regions (Fig.~\ref{fig2}a--c). We obtained the 1-Hz residual $\delta C/N_0$ by subtracting the arc-fitted background from the receiver-reported $C/N_0$ (Extended Data Fig.~\ref{edfig:cn0-detrending-op38-e21}). The global GPS L1/Galileo E1 map shows the 95th percentile ($P_{95}$) of $|\delta C/N_0|$ within each $5^\circ\times5^\circ$ tangent-point cell over the observing campaign (Fig.~\ref{fig2}a). {Using $P_{95}$ isolates large $C/N_0$ fluctuations while excluding severe outliers (Extended Data Fig.~\ref{edfig:cn0-quality-background}).} The polar maps show the individual 1-Hz $|\delta C/N_0|$ values along the tangent tracks (Fig.~\ref{fig2}b,c).
}

% R3_RESULTS_FIG2_P02
{
The rate-of-TEC index (ROTI) was calculated from dual-frequency carrier-phase observations over 1-min windows and mapped using $P_{95}$ within the same spatial cells (Fig.~\ref{fig2}d). Elevated ROTI extended across a broader latitude range than the large amplitude fluctuations.
}

The original dual-frequency $C/N_0$ measurements showed individual fades exceeding 10~dB (Fig.~\ref{fig2}e,f). Fluctuations spanned approximately 150--650~km in tangent height in the northern GPS G01 crossing and 150--250~km in the southern Galileo E21 crossing.

{
For both GPS and Galileo, polar spectra showed greater power above 0.1~Hz than equatorial and mid-latitude spectra in both carrier-frequency groups (Fig.~\ref{fig2}g,j). The power spectral densities (PSDs) are calculated from continuous 1-Hz $\delta C/N_0$ segments along individual satellite--signal tracks and obtained median spectra for each signal and geomagnetic-latitude band.
}

{
We measured the disturbance width $W=P_{95}(\delta C/N_0)-P_{05}(\delta C/N_0)$ in each 1-min window. In the paired observations, the strongest polar responses generally had larger widths at L5/E5a than at L1/E1 (Fig.~\ref{fig2}h). \revSJTU{This is consistent with the wavelength dependence of amplitude scintillation: under weak-scattering conditions, the longer-wavelength L5/E5a signals (1176.45~MHz) are expected to exhibit stronger amplitude scintillation than L1/E1 (1575.42~MHz) \cite{YehLiu1982Scintillations,Aguiar2025ScintillationFrequency}.} 
}

\revSJTU{At tangent heights above 1{,}000~km, background $C/N_0$ deviations typically spanned approximately 0.5~dB and reflected signal noise in the onboard $C/N_0$ estimates (Extended Data Fig.~\ref{edfig:cn0-quality-background}e,f). We therefore classified 1-min windows with $W\geq1.5$~dB as ionospheric disturbance events, using a threshold three times this background range.} Across GPS L1/L5 and Galileo E1/E5a, 81\% of the event windows came from the polar regions. {Within the polar observations, 30\% of valid windows met this criterion (207/680), compared with 4.1\% at equatorial latitudes (21/515) and 2.6\% at mid-latitudes (26/1{,}001).}
Polar event widths reached approximately 10~dB (Fig.~\ref{fig2}k). The polar event fraction was {46\%} at tangent heights of 300--600~km and 15\% at 600--1{,}000~km (Fig.~\ref{fig2}i).

%%%%%%%%%%%%%%%%%%%%%%%%%%%%%%%%%%Fig.3%%%%%%%%%%%%%%%%%%%%%%%%%%%%%%%%%%%%%%%%%%%%%%%%%%%%
{
To investigate the polar enhancement observed from the Moon, we first considered whether it coincided with heightened geomagnetic activity as the March equinox approached. \revSJTU{We examined the planetary K-index (Kp), the polar cap activity index for Northern Hemisphere (PCN) and the auroral electrojet indices (AE, AU and AL) to assess geomagnetic storm, polar-cap and auroral activity \cite{Kauristie2017GeomagneticIndices}.} Despite intermittent increases in polar-cap and auroral activity, pronounced lunar fluctuations also occurred during relatively quiet geomagnetic intervals (Extended Data Fig.~\ref{edfig:solarwind-lugre-context}).
}

{
{Because received $C/N_0$ also varies with propagation loss, Earth blockage, antenna gain and neutral-atmospheric effects \cite{Parker2026Navigation}, we examined whether these factors could contribute to the observed amplitude fluctuations. Range-dependent free-space loss and transmitter antenna gain vary relatively slowly along the illustrated arc and are unlikely to explain the rapid, deep fades (Extended Data Fig.~\ref{edfig:cn0-detrending-op38-e21}). At the selected tangent heights of 50--1{,}000~km, the rays remain clear of the solid Earth and dense lower atmosphere, limiting blockage and neutral-atmospheric attenuation. We therefore compared transmitter off-boresight angles, $\alpha$, with GPS and Galileo antenna patterns to test whether gain changes near main-lobe or sidelobe edges could explain the deep fades \cite{Marquis2016GPSBlockIIR,Menzione2024GalileoAntenna}. The transmitter off-boresight angles for observations at tangent heights of 50--1{,}000~km corresponded to relatively high-gain portions of the main lobes, making transitions into sidelobes unlikely (Extended Data Fig.~\ref{edfig:main-lobe-antenna}). Together with the height-dependent dual-frequency fluctuations (Fig.~\ref{fig2}e,f) and the smaller \revSJTU{background $C/N_0$ deviations} (Extended Data Fig.~\ref{edfig:cn0-quality-background}e,f), these checks support ionospheric irregularities as the source of these amplitude fluctuations.}
}

% R3_RESULTS_FIG3_CONTEXT
{We next examined whether near-Earth observations showed the same polar enhancement during the lunar observing intervals. The comparison used receiver-reported 1-Hz $C/N_0$ from 182 continuously operating GNSS stations, including CORS/IGS sites, and amplitude observations derived from PlanetiQ, Spire, TerraSAR-X and MetOp radio-occultation products \cite{Chang2025PlanetiQSpireRO}. Within each magnetic-latitude band, near-Earth observations were restricted to the UTC intervals sampled by LuGRE (Extended Data Fig.~\ref{edfig:observing-context}).}

\begin{figure}[tbp]
\centering
\includegraphics[width=\textwidth]{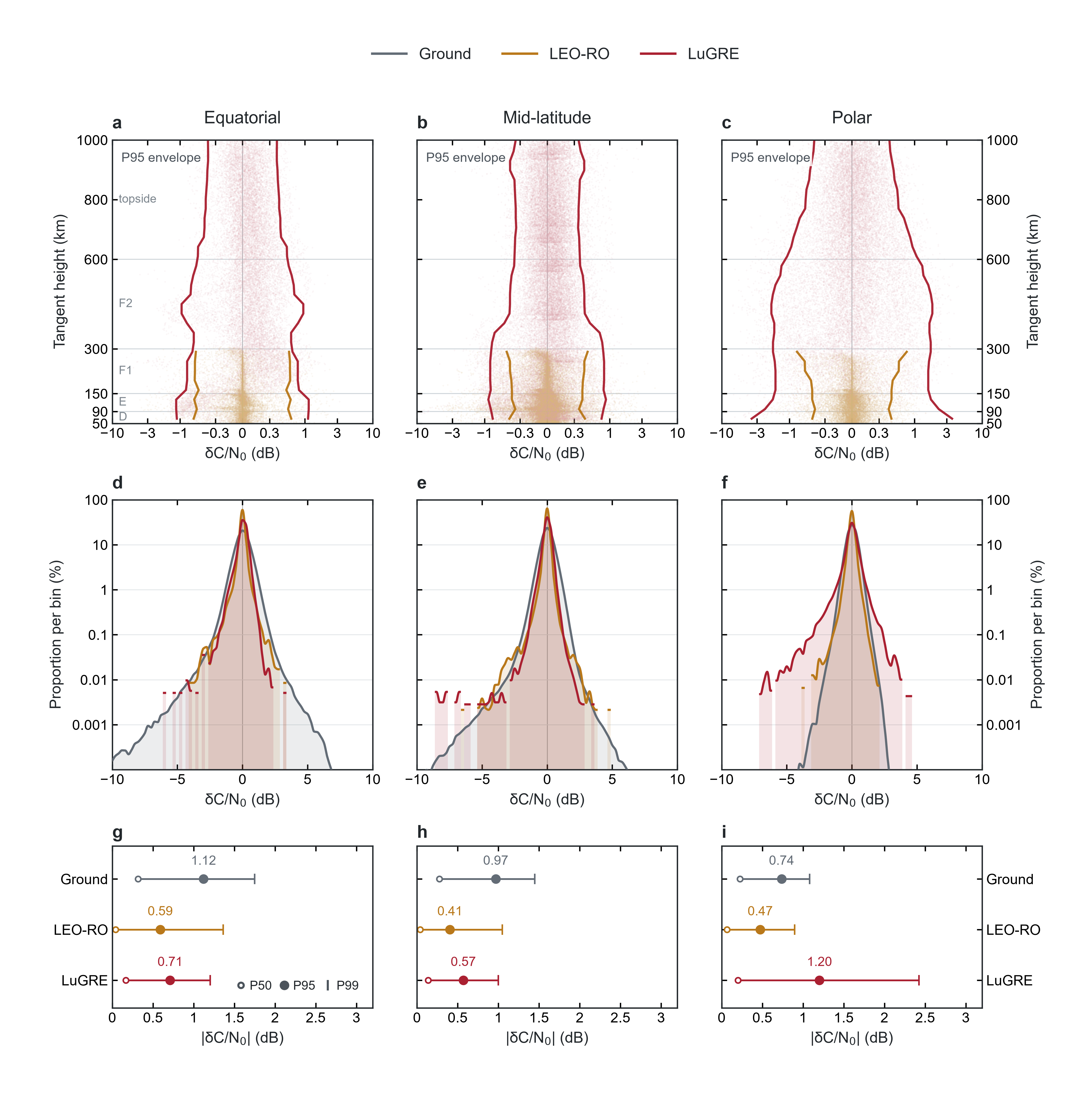}
\small
\caption{\textbf{Lunar observations extend height coverage and reveal polar amplitude enhancement.}
\textbf{a--c}, Detrended 1-Hz GPS L1/Galileo E1 fluctuations, $\delta C/N_0$, versus tangent height, with envelopes defined by $P_{95}$ of $|\delta C/N_0|$. The LEO-RO profiles cover tangent heights below approximately 300~km, whereas lunar observations extend into the topside ionosphere. Polar lunar fluctuations are stronger than those at equatorial and mid-latitudes.
\textbf{d--f}, Distributions of the 1-Hz fluctuations from Ground, LEO-RO and LuGRE. Columns show equatorial, mid-latitude and polar observations.
\textbf{g--i}, The 50th, 95th and 99th percentiles of $|\delta C/N_0|$ summarize the distributions in \textbf{d--f}. The upper percentiles are largest in the polar band for LuGRE and in the equatorial band for Ground and LEO-RO.}

\label{fig3}
\end{figure}

% R3_RESULTS_FIG3_NEW_P1
{
{We compared 1-Hz GPS L1/Galileo E1 fluctuations across the three platforms. Slow backgrounds were removed using {fourth-order polynomial fits to the Ground and LuGRE arcs} and rolling medians for the shorter LEO-RO profiles \cite{Cornman2012ROTurbulence} (Methods). Lunar and near-Earth observations differed in both height coverage and latitude dependence (Fig.~\ref{fig3}).}
}

% R3_RESULTS_FIG3_NEW_P2
Lunar reception extended the height range over which irregularity-related amplitude fluctuations were observed (Fig.~\ref{fig3}a--c). The sampled LEO-RO profiles covered tangent heights below approximately 300~km, whereas LuGRE reached into the topside ionosphere. Polar lunar fluctuations remained elevated through the F region and lower topside, exceeding the polar LEO-RO envelope over the overlapping height range before decreasing at greater heights.

% R3_RESULTS_FIG3_NEW_P3
The fluctuation distributions revealed a distinct latitude ordering (Fig.~\ref{fig3}d--i). LuGRE's polar distribution was broader than its equatorial and mid-latitude distributions, with a $P_{95}$ of $|\delta C/N_0|$ of {1.20}~dB compared with {0.71} and {0.57}~dB, respectively. Ground and LEO-RO instead had their largest upper-percentile fluctuations at equatorial latitudes. The lunar polar enhancement thus contrasts with the equatorial emphasis of these near-Earth observations.

% R3_RESULTS_FIG3_NEW_P4
{
This latitude contrast occurred under predominantly sunlit lunar sampling (Extended Data Fig.~\ref{edfig:local-time-illumination}). Nearly all equatorial lunar windows fell between 06 and 11 local time, with sparse coverage after sunset; polar lunar observations were also predominantly sunlit. Ground observations covered local time more broadly, whereas LEO-RO measurements clustered in morning and evening sectors. Equatorial plasma bubbles (EPBs) and their associated irregularities commonly develop \revSJTU{after local sunset at geomagnetic equator and ionospheric altitudes} \cite{Spogli2023EPB}. Limited sampling of these post-sunset irregularities may therefore contribute to the weaker equatorial amplitude fluctuations observed from the Moon.
}

%%%%%%%%%%%%%%%%%%%%%%%%%%%%%%%%%%Fig.4%%%%%%%%%%%%%%%%%%%%%%%%%%%%%%%%%%%%%%%%%%%%%%%%%%%%

\begin{figure}[tbp]
\centering
\includegraphics[width=\textwidth]{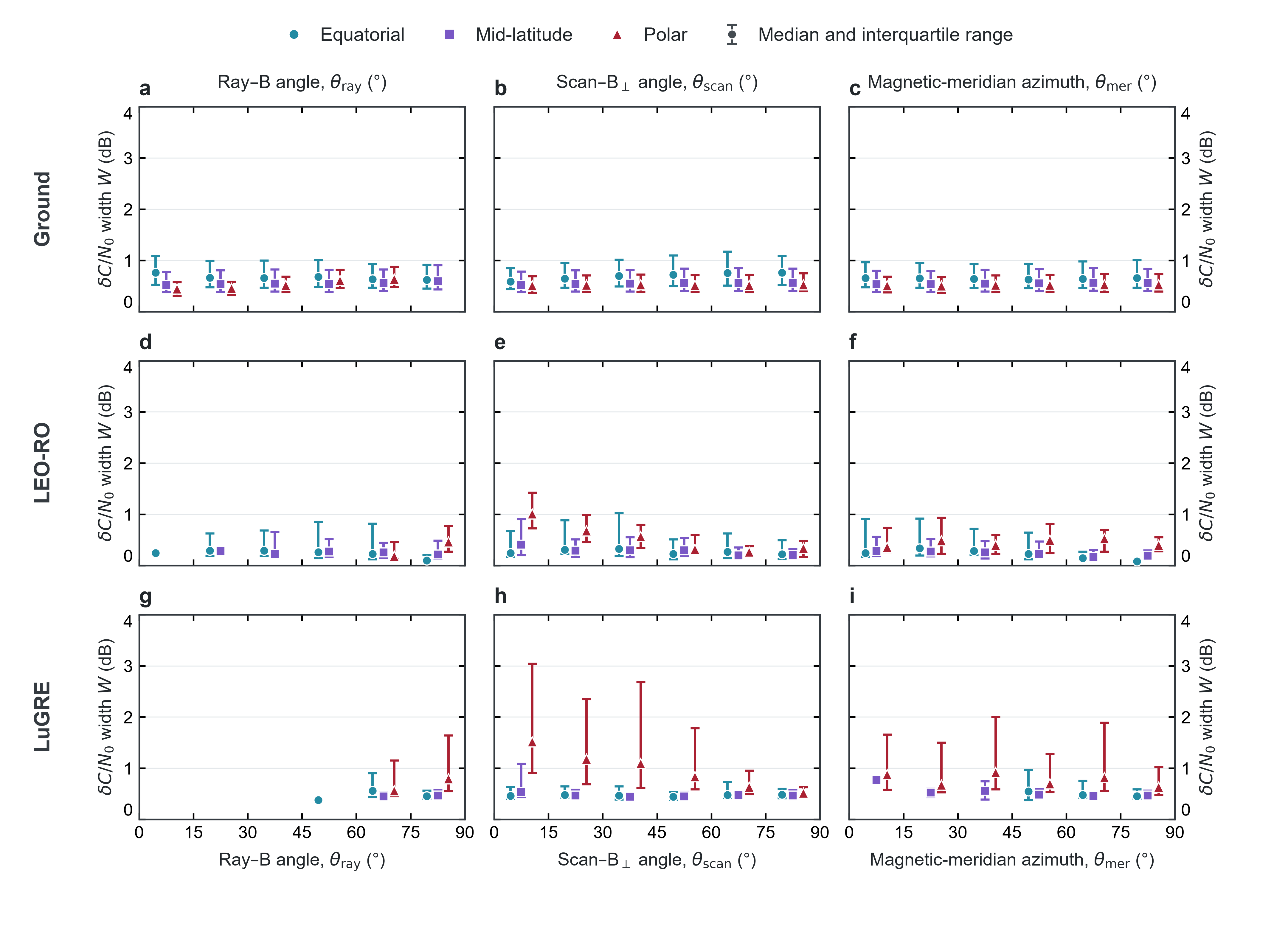}
\small
\caption{\textbf{Polar lunar fluctuation widths decrease as scanning turns across the projected magnetic field.}
Rows show Ground (\textbf{a--c}), LEO-RO (\textbf{d--f}) and LuGRE (\textbf{g--i}). Columns show the ray--field angle $\theta_{\rm ray}$, scan--projected-field angle $\theta_{\rm scan}$ and magnetic-meridian azimuth $\theta_{\rm mer}$.
Ground widths show comparatively little angular variation across the three latitude bands. In LEO-RO, the clearest ordered change is a gradual decrease in polar widths with increasing $\theta_{\rm scan}$.
The lunar polar response has the strongest scan-angle contrast: widths are greater when scanning follows the projected magnetic field and decrease towards perpendicular scanning (\textbf{h}). Equatorial and mid-latitude lunar widths remain comparatively stable. Within the sampled near-perpendicular ray--field geometries, polar lunar widths increase towards perpendicular ray orientations (\textbf{g}), while their variation with magnetic-meridian azimuth is less ordered (\textbf{i}).
The possible roles of propagation scale and directional sampling are sketched in Fig.~\ref{fig1}b--f.}

\label{fig4}
\end{figure}

% R3_RESULTS_FIG4_NEW_P1
The lunar polar enhancement varied most clearly with scanning relative to the magnetic field (Fig.~\ref{fig4}). Fluctuation widths were greatest when scanning followed the projected magnetic field and decreased as the scan direction approached perpendicularity, with an approximately threefold contrast between the two orientations (Fig.~\ref{fig4}h). Equatorial and mid-latitude lunar widths remained comparatively stable. Polar LEO-RO widths showed a similar overall decline, while Ground widths varied little with scan direction.

% R3_RESULTS_FIG4_NEW_P2
Ray orientation showed a different relationship. Within the sampled near-perpendicular geometries, polar lunar widths increased as rays approached perpendicularity to the magnetic field (Fig.~\ref{fig4}g). All three lunar latitude bands occupied similar ray--field angle ranges (Extended Data Fig.~\ref{edfig:geometry-angles}), yet their amplitude responses differed. Polar widths varied less systematically with magnetic-meridian azimuth (Fig.~\ref{fig4}i), leaving transverse scan orientation as the clearest angular association.

% R3_RESULTS_FIG4_NEW_P3
{
At lunar distance, the larger Fresnel scale and \revSJTU{the smaller scan speed} yield longer geometric crossing times than in LEO-RO (Extended Data Fig.~\ref{edfig:scan-speed-fresnel}). Under suitable irregularity scales and orientations, this propagation geometry could strengthen the amplitude response to polar structures. The observed polar scan-angle association may reflect a sensitivity to ray and scan orientations relative to field-aligned irregularities, as sketched in Fig.~\ref{fig1}b--f. The geometric crossing times do not include irregularity drift, which would change both the speed and direction of sampling relative to the plasma.
}

%%%%%%%%%%%%%%%%%%%%%%%%%%%%%%%%%%Fig.5%%%%%%%%%%%%%%%%%%%%%%%%%%%%%%%%%%%%%%%%%%%%%%%%%%%%
% R3_RESULTS_FIG5_P01
{
Pronounced lunar amplitude fluctuations are supported by independent evidence of polar ionospheric structure (Fig.~\ref{fig5}). The LuGRE ray paths cross polar sectors containing SuperDARN ionospheric backscatter and enhanced regional ground ROTI (Fig.~\ref{fig5}a,b,h,j). The radar echoes identify field-aligned irregularities, and the ROTI enhancement shows variability in integrated electron content. Regional electron-density profiles additionally show F-region structure and enhancement (Fig.~\ref{fig5}g,i).
}

% R3_RESULTS_FIG5_P02
{
In both regions, ground $S_4$ indicates predominantly weak amplitude scintillation (Fig.~\ref{fig5}a,b). ROTI and $S_4$ characterize electron-content disturbances and amplitude scintillation, respectively. The contrast shows that strong lunar amplitude responses can coexist with marked electron-content variability and weak ground amplitude scintillation. The comparison across latitude bands likewise highlights LuGRE's polar amplitude enhancement alongside near-Earth high-rate amplitude and phase scintillation indices (Extended Data Fig.~\ref{edfig:ROTI_S4}).
}

% R3_RESULTS_FIG5_P03
Compared with the regional LEO-RO profiles, LuGRE shows larger 1-Hz amplitude fluctuations across a wider range of tangent heights (Fig.~\ref{fig5}c--f). Together, the two lunar cases show pronounced fluctuations from approximately 100 to 600~km in tangent height. The corresponding MetOp-B and PlanetiQ fluctuations are smaller and concentrated at the lower tangent heights sampled by the occultations.

% Retain Fig.5 numbering after the completed Fig.3--4 draft.
\setcounter{figure}{4}
\begin{figure}[tbp]
\centering
\includegraphics[width=\textwidth]{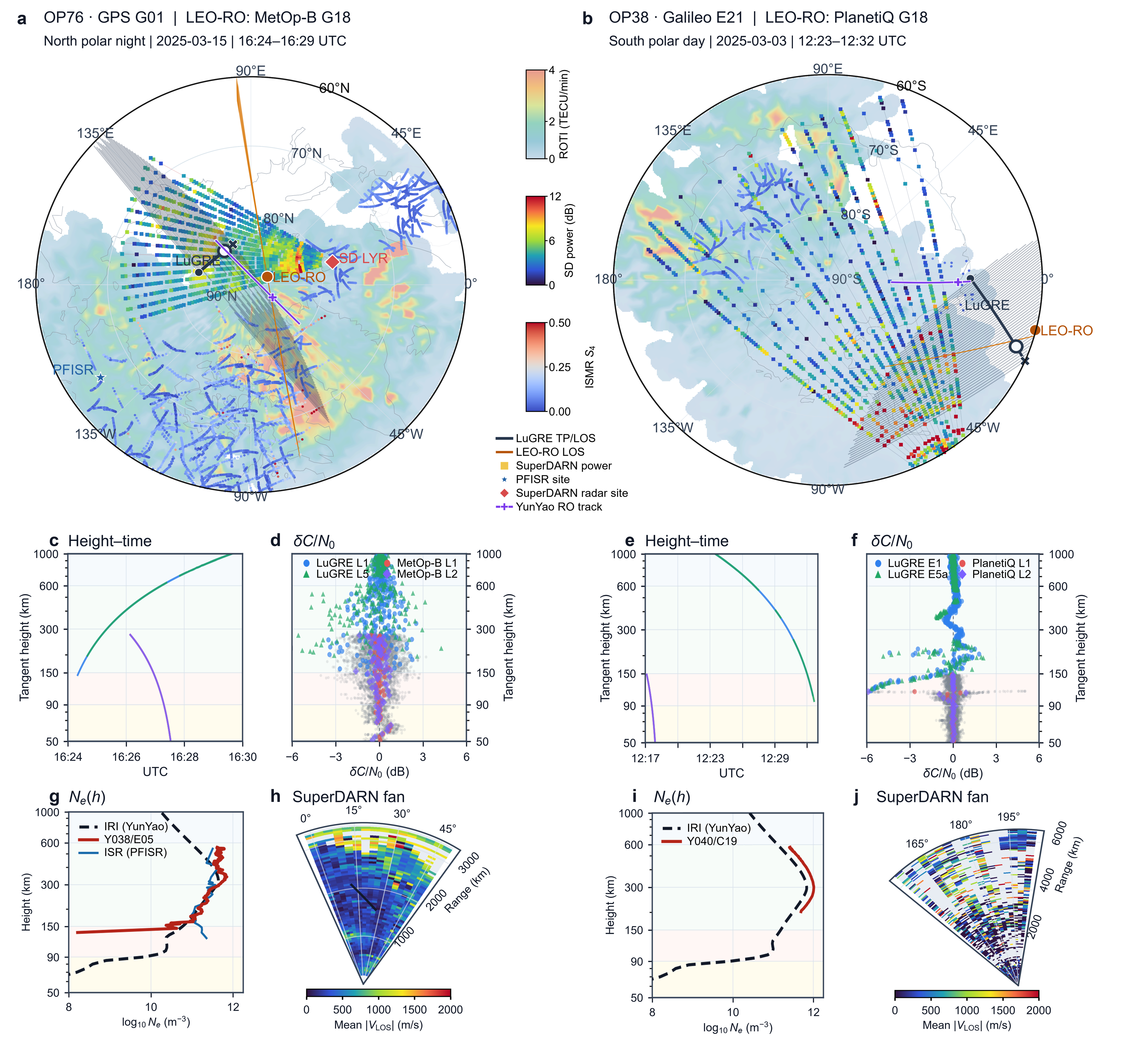}
\small
\caption{\textbf{Independent observations reveal structured plasma around polar lunar GNSS crossings.}
\textbf{a,b}, Northern OP76 GPS G01 (15 March 2025) and southern OP38 Galileo E21 (3 March 2025) paths, with regional ground ROTI, ground $S_4$, SuperDARN and LEO-RO context. ROTI backgrounds summarize 15:00--18:00 UTC in the north and 11:00--15:00 UTC in the south.
\textbf{c,e}, Tangent height versus UTC for LuGRE and regional LEO-RO occultations.
\textbf{d,f}, Dual-frequency LuGRE and 1-Hz LEO-RO amplitude fluctuations. Grey points show high-rate LEO-RO observations, which resolve larger short-timescale fluctuations than the 1-Hz measurements. 
\textbf{g,i}, YunYao electron-density profiles show F-region structure and enhanced peak densities relative to the International Reference Ionosphere (IRI) background \cite{Bilitza2022IRI}. Profiles are Y038/E05 (north, approximately 16:00 UTC) and Y040/C19 (south, approximately 12:17 UTC); contemporaneous PFISR measurements provide an independent regional density profile in the north.
\textbf{h,j}, SuperDARN mean absolute LOS velocity, showing the magnitude of ionospheric plasma motion along the radar beam.}
\label{fig5}
\end{figure}

\section{Discussion}\label{sec3}
% R3_DISCUSSION_P01
{
This work presents the first systematic characterization of terrestrial ionospheric irregularities using GNSS amplitude observations from the lunar surface. At lunar distance, the Fresnel scale approaches its finite-source limit, extending the study of irregularity-driven signal fluctuations to propagation conditions beyond those of ground and LEO observations \cite{Rino2011GPSScintillation}. The pronounced polar amplitude enhancement reveals a distinctive latitude dependence of the irregularity response from this distant viewpoint. The polar enhancement further suggests that amplitude fluctuations in this regime may be particularly sensitive to ray--field alignment and transverse scanning across field-aligned irregularities.
}

% R3_DISCUSSION_P02
{
Local-time sampling limits how broadly the observed latitude ordering can be generalized. Equatorial observations were concentrated in the morning, with sparse coverage of post-sunset conditions \revSJTU{likely} associated with equatorial plasma bubbles \cite{Spogli2023EPB}. Future missions should fill this gap and compare latitude bands under comparable local-time and illumination conditions. If polar enhancement persists, ray and scan orientations relative to field-aligned irregularities could favour stronger amplitude modulation along lunar-distance paths \cite{Rino1979WeakScatter}.
}

% R3_DISCUSSION_P03
{
High-rate amplitude and phase observations are essential for resolving the rapid fluctuations produced by ionospheric irregularities. The conventional amplitude and phase scintillation indices, $S_4$ and $\sigma_\phi$, would provide a fuller characterization of scintillation than the present 1-Hz $\delta C/N_0$ proxy \cite{Spogli2009GlobalScintillation,Prikryl2015PolarScintillation}. Extending lunar observations to these measurements would reveal short-timescale variability unresolved here and enable closer tests of how irregularity scales and anisotropy shape scintillation in the lunar-distance propagation regime.
}

% R3_DISCUSSION_P04
NASA Artemis program\cite{NASAArtemisII2026}, ESA Moonlight infrastructure\cite{ESAMoonlight2024}, Chinese Chang'e missions towards the International Lunar Research Station\cite{CNSAChange82025}, and the joint India--Japan LUPEX mission\cite{JAXALUPEX2023} are expanding opportunities for sustained lunar exploration. Future lunar-surface receivers could establish a long-term platform for observing Earth's ionosphere, revealing irregularity responses and propagation phenomena under lunar-distance geometries. Repeated observations across seasons and different solar and geomagnetic conditions could open a new direction in lunar-based remote sensing of Earth's ionosphere. Understanding the associated amplitude fading would also inform future lunar mission engineering and receiver design.

{
\section{Methods}\label{sec4}

\subsection{GNSS observations and propagation geometry}

We analysed nearly 93~h of receiver-reported 1-Hz $C/N_0$ and carrier observations from LuGRE's lunar-surface operations on 3--16 March 2025, {including} GPS L1/L5 and Galileo E1/E5a \cite{LuGREData2025,Parker2026Navigation}. Receiver positions were obtained from the NAIF CLPS SPICE trajectory for \texttt{BGM1\_LUGRE} relative to Earth; {precise GNSS ephemerides} were interpolated to the observation epochs \cite{NAIFCLPSSPICE}. Both were expressed in the Earth-centred frame.

{For the straight-ray geometry used here, the tangent point was defined as the point on the transmitter--receiver line of sight (LOS) closest to Earth's centre \cite{Wu2009IROAsymmetry}. Its WGS84 geodetic height defined the tangent height.} For receiver and transmitter positions $\mathbf r_{\rm rx}$ and $\mathbf r_{\rm tx}$, we defined {the LOS vector $\mathbf d=\mathbf r_{\rm tx}-\mathbf r_{\rm rx}$} and
\begin{equation}
s_{\rm tan}=-\dfrac{\mathbf r_{\rm rx}\cdot\mathbf d}{|\mathbf d|^2},
\qquad
\mathbf r_{\rm tan}=\mathbf r_{\rm rx}+s_{\rm tan}\mathbf d .
\label{eq:r3-tangent}
\end{equation}
We retained $0\leq s_{\rm tan}\leq1$ and transformed the tangent point to Earth-fixed coordinates. Observations with WGS84 tangent heights of 50--1{,}000~km were selected to examine ionospheric irregularities. The same calculation and height range were applied to LEO-RO using the GNSS occultation products. 

{Links to ground monitoring stations} were located by their ionospheric pierce points (IPPs), calculated at the intersection of each ray with a 350-km ionospheric thin shell \cite{Smith2024PhonesIonosphere,Spogli2009GlobalScintillation}. 
For latitude-band classification, centred-dipole magnetic latitude $\lambda_m$ at each tangent point or IPP was calculated relative to a fixed reference pole at $80.65^\circ$N, $72.68^\circ$W for the equatorial ($|\lambda_m|<20^\circ$), mid-latitude ($20^\circ\leq|\lambda_m|<60^\circ$) and polar ($|\lambda_m|\geq60^\circ$) bands \cite{Zolesi2014IonosphericPrediction}. Individual epochs used their own reference positions; track and profile labels used the median geomagnetic latitude. 

{
Height bands followed the typical daytime ionospheric structure: D (50--90~km), E (90--150~km), F1 (150--300~km), F2 (300--600~km) and topside (600--1{,}000~km) \cite{Zolesi2014IonosphericPrediction,Rishbeth2002Ionosphere}. At night, the D and E regions weaken and the F region is usually dominated by the F2 layer.
}

Solar geometry was evaluated at the 350-km IPP for Ground and at the tangent point for LuGRE and LEO-RO. The solar zenith angle was
\begin{equation}
\mathrm{SZA}(t)=\arccos\!\left[
\hat{\mathbf n}(t)\cdot\hat{\mathbf s}_{\odot}(t)
\right],
\label{eq:r3-solar-zenith}
\end{equation}
where $\hat{\mathbf n}$ is the local upward unit vector normal to the WGS84 ellipsoid and $\hat{\mathbf s}_{\odot}$ points towards the solar centre. Local solar time, in hours, was $\mathrm{LT}=[12+(\ell-\ell_{\rm ss})/15^\circ]\bmod24$, where $\ell$ and $\ell_{\rm ss}$ are the east-positive longitudes of the reference point and subsolar point.

Illumination was evaluated at each ionospheric reference point at every observed epoch. A point was classified as \emph{Sunlit} when its ray towards the solar centre did not intersect the solid WGS84 ellipsoid, and as \emph{Earth shadow} when it did. The local-time and illumination coverage is shown in Extended Data Fig.~\ref{edfig:local-time-illumination}.

\subsection{Amplitude and phase-response processing of ionospheric irregularities}
The amplitude scintillation index $S_4$ and phase scintillation index $\sigma_\phi$ are widely used to characterize ionospheric irregularities. They are calculated from high-rate (usually 50 or 100~Hz) intensity and carrier-phase measurements, respectively. For example, ground-based $S_4$ and $\sigma_\phi$ products are available from ionospheric scintillation monitoring networks \cite{Spogli2009GlobalScintillation,CardenasOToole2024PolarPatches}. LEO-RO missions also provide high-rate SNR and excess-phase observations for deriving these indices. In this study, we calculated $S_4$ and $\sigma_\phi$ in 10-s sliding windows \cite{Chang2025PlanetiQSpireRO}.

More extensive ground-based geodetic networks, including CORS stations, provide dual-frequency carrier-phase observations at 1- or 30-s intervals for calculating ROTI (rate-of-TEC index), another widely used measure of ionospheric irregularities \cite{Mrak2023GroundInfrastructure,Spogli2023EPB}. 

In this work, the available LuGRE amplitude and carrier observations were sampled at 1~Hz. We therefore used detrended $C/N_0$ fluctuations to characterize amplitude response on each satellite signal, supplemented by ROTI from the dual-frequency carrier measurements \cite{Cesaroni2026ESSOAr}.

{For each LuGRE satellite--signal propagation geometry, we reconstructed complete observation arcs by joining tracking segments separated by gaps of up to 2~h, including Earth-occultation gaps. A single fourth-order polynomial was fitted to the available measurements over each reconstructed arc (Extended Data Fig.~\ref{edfig:cn0-detrending-op38-e21}),}
\begin{equation}
{\widehat{C/N}}_0^k(t)=\sum_{n=0}^{4}a_n^k[x^k(t)]^n ,
\label{eq:cn0-poly4-baseline}
\end{equation}
where $k$ identifies {a reconstructed satellite--signal arc}, $x^k(t)$ is time centred and scaled within that arc, and $a_n^k$ are its fitted coefficients.  

Amplitude fluctuations were expressed as
\begin{equation}
\delta C/N_0^k(t_i)=C/N_0^k(t_i)-{\widehat{C/N}}_0^k(t_i).
\label{eq:cn0-residual}
\end{equation}

For a fair comparison, 1 Hz $\delta C/N_0$ is also calculated for ground stations and LEO-RO. 
The same {fourth-order polynomial fit} was applied to each continuous ground GNSS receiver--satellite--signal arc. Because such long durations help constrain the slowly varying background in LuGRE and ground stations.

LEO-RO observations covered shorter profiles as the rapidly moving receiver scanned the Earth's limb. For these profiles, high-rate SNR was reduced to its median within each UTC second to form an amplitude proxy. The short profiles were then detrended using a centred rolling-median background, following an earlier application to GNSS occultation amplitudes \cite{Cornman2012ROTurbulence}.

All ground, LEO-RO and LuGRE observations underwent quality screening. Ground measurements were retained at elevation angles $\geq30^\circ$ and $C/N_0\geq30$~dB-Hz. Ground intervals with confirmed direction-dependent slow backgrounds were excluded in full from the affected receiver--satellite--signal observations. For LEO-RO, the median of the amplitude residuals was checked for persistent background offsets. 

{For LuGRE, the LuGRE Navigation Analysis and Reporting Tool (LuNART) was used to inspect tracking products \cite{NavsasPolitoLuNART}. Each satellite--signal arc was reconstructed and fitted independently, with simultaneous dual-frequency observations used to assess background consistency. Arcs with too few observations for a fourth-order polynomial fit were excluded from amplitude statistics.}

{Before calculating lunar amplitude statistics, we excluded measurements with $C/N_0<25$~dB-Hz to reduce fluctuations associated with unstable tracking. The remaining $\delta C/N_0$ observations were grouped into 60-s windows. Valid windows required $|\mathrm{median}(\delta C/N_0)|\leq1$~dB to limit residual background offsets, and at least 30 valid 1-Hz observations ($N_{\rm valid}\geq30$) to avoid unstable fluctuation estimates from sparse sampling (Extended Data Fig.~\ref{edfig:cn0-quality-background}a--d).}

{Within each valid 1-min window, the fluctuation width $W=P_{95}(\delta C/N_0)-P_{05}(\delta C/N_0)$ characterized the fluctuation range. For $|\delta C/N_0|$, we used $P_{95}$ to characterize large fluctuation magnitudes. These quantile-based summaries reduced the influence of extreme values. \revSJTU{At tangent heights above 1{,}000~km, background $C/N_0$ fluctuations typically span approximately 0.5~dB and probably reflect noise-induced deviations in the onboard $C/N_0$ estimates (Extended Data Fig.~\ref{edfig:cn0-quality-background}e,f). We defined windows with $W\geq1.5$~dB as ionospheric disturbance events, using a threshold 3 times this background range.} The higher polar event fraction persisted when the threshold was reduced to 2 or 1.5 times \revSJTU{this background range}.}

\subsection{Ray path scan geometry}

We compared propagation distances, scan speeds and magnetic-field orientations at the ionospheric reference point $\mathbf r_{\rm p}$: the tangent point for LuGRE and LEO-RO, and the 350-km IPP for ground links. Positions and velocities were expressed in an Earth-fixed frame. {The unit ray direction was $\hat{\mathbf k}=\mathbf d/|\mathbf d|$.}

{
Using the phase screen model at the reference point, the screen-to-transmitter and screen-to-receiver distances were $D_{\rm tx}=|\mathbf r_{\rm tx}-\mathbf r_{\rm p}|$ and $D_{\rm rx}=|\mathbf r_{\rm rx}-\mathbf r_{\rm p}|$. The first Fresnel-zone radius in Fig.\ref{fig1} was
\begin{equation}
r_1=\sqrt{\lambda\frac{D_{\rm tx}D_{\rm rx}}{D_{\rm tx}+D_{\rm rx}}},
\label{eq:r3-fresnel}
\end{equation}
where $\lambda$ is the signal wavelength. As $D_{\rm rx}$ increases, $r_1$ approaches the finite-source limit $\sqrt{\lambda D_{\rm tx}}$ \cite{Morton2020IonosphericEffects,Rino1979WeakScatter,Rino2011GPSScintillation}.
}

The motion of the tangent point or IPP was calculated by finite differences of its positions along each continuous trajectory. Its velocity and speed were
\begin{equation}
\mathbf v_{\rm p}=\dfrac{\mathrm d\mathbf r_{\rm p}}{\mathrm dt},
\qquad
v_{\rm p}=|\mathbf v_{\rm p}|.
\label{eq:r3-reference-point-speed}
\end{equation}
The component perpendicular to the ray describes how rapidly the link scans across the transverse plane.

For the propagation and scan angle calculations, the local magnetic-field vector $\mathbf B$ was evaluated using the 14th-generation International Geomagnetic Reference Field (IGRF-14) at each reference position and observation time. The transverse scan velocity and magnetic-field projection were
\begin{equation}
\mathbf v_{\rm scan}=\mathbf v_{\rm p}-(\mathbf v_{\rm p}\cdot\hat{\mathbf k})\hat{\mathbf k},
\qquad
\mathbf B_\perp=\mathbf B-(\mathbf B\cdot\hat{\mathbf k})\hat{\mathbf k},
\label{eq:r3-transverse-projections}
\end{equation}
with scan speed $v_{\rm scan}=|\mathbf v_{\rm scan}|$. The ray--field and scan--field angles were
\begin{equation}
\theta_{\rm ray}=\arccos\dfrac{|\hat{\mathbf k}\cdot\mathbf B|}{|\mathbf B|},
\qquad
\theta_{\rm scan}=\arccos\dfrac{|\mathbf v_{\rm scan}\cdot\mathbf B_\perp|}
{|\mathbf v_{\rm scan}|\,|\mathbf B_\perp|}.
\label{eq:r3-geometry-angles}
\end{equation}
The first angle describes propagation relative to the local field; the second describes scanning relative to its projection onto the ray-normal plane. Both range from $0^\circ$ for parallel orientations to $90^\circ$ for perpendicular orientations.

The azimuth relative to the magnetic meridian was calculated from the local horizontal projections of the ray and magnetic field. Using the local upward unit normal $\hat{\mathbf n}$, these projections and their acute angle were
\begin{equation}
\begin{array}{c}
\mathbf k_{\rm h}=\hat{\mathbf k}-(\hat{\mathbf k}\cdot\hat{\mathbf n})\hat{\mathbf n},
\qquad
\mathbf B_{\rm h}=\mathbf B-(\mathbf B\cdot\hat{\mathbf n})\hat{\mathbf n},
\\[6pt]
\displaystyle
\theta_{\rm mer}=\arccos\dfrac{|\mathbf k_{\rm h}\cdot\mathbf B_{\rm h}|}
{|\mathbf k_{\rm h}|\,|\mathbf B_{\rm h}|}.
\end{array}
\label{eq:r3-magnetic-meridian-angle}
\end{equation}
Here $0^\circ$ indicates a ray azimuth along the local magnetic meridian and $90^\circ$ indicates a perpendicular azimuth.

{
We combined propagation scale and scan speed through the geometric Fresnel crossing time,
\begin{equation}
\tau_{\rm F,geom}=\dfrac{r_1}{v_{\rm scan}},\qquad v_{\rm scan}>0.
\label{eq:r3-fresnel-crossing-time}
\end{equation}
This is the time required to scan one first-zone radius. It provides a geometric time scale for comparing the ground's, LEO's, and Lunar surface's disturbances; the relative motion through plasma structures also depends on plasma drift.
}
}

\backmatter

\section*{Acknowledgements}

We thank Qinghe Zhang (State Key Laboratory of Solar Activity and Space Weather, National Space Science Center, Chinese Academy of Sciences) for guidance on this work, and Haiyang Fu (School of Information Science and Technology, Fudan University) and Dun Liu (China Research Institute of Radiowave Propagation) for helpful discussions and advice. We thank Tianjin Yunyao Aerospace Technology Co., Ltd. for providing access, upon request, to YunYao ionPrf electron-density profiles for the March 2025 LuGRE event intervals.

\section*{AI assistance statement}
ChatGPT and Codex (OpenAI) were used during manuscript preparation to assist with language editing, code review and figure-script refinement. All scientific analyses, interpretations, figures and final text were checked and approved by the authors.

\begingroup

% Keep the Data availability heading with its paragraph.
\par
\ifdim\dimexpr\pagegoal-\pagetotal\relax<5\baselineskip
  \newpage
\fi
\section*{Data availability}

% R3_DATA_GROUND_SPLIT: 1-Hz amplitude, polar TEC/ROTI, high-rate scintillation.
\begingroup
{Processed data and figure source data are available at \url{https://doi.org/10.5281/zenodo.21672260}.} Raw LuGRE observations are available from the NASA--ASI mission archive (\url{https://doi.org/10.5281/zenodo.16411687}), and trajectory kernels from the NASA/JPL NAIF CLPS SPICE archive (\url{https://naif.jpl.nasa.gov/pub/naif/pds/pds4/clps/clps_spice/}) \cite{LuGREData2025,NAIFCLPSSPICE}. {Precise GNSS ephemerides in SP3 format} are available through IGS (\url{https://igs.org/data-access/}).

Ground 1-Hz $C/N_0$ observations from 182 GNSS stations on 3--4 and 14--16 March 2025 were used for the amplitude comparison. These data are available from EarthScope (\url{https://data.unavco.org/archive/gnss/}), NASA CDDIS (\url{https://cddis.nasa.gov/archive/gnss/}), BKG (\url{https://igs.bkg.bund.de/root_ftp/}), IGN (\url{ftp://igs.ign.fr/pub/igs/data/highrate/}), Geoscience Australia (\url{https://data.gnss.ga.gov.au/}), the NRCan Canadian Active Control System (\url{https://cacsa.nrcan.gc.ca/gps/data/hrdata/}), the Pacific Northwest Geodetic Array (\url{https://www.geodesy.org/pub/data/}), the Florida Permanent Reference Network (\url{https://www.myfloridagps.com/ftp/Rinex3/01/}), the US National Park Service GNSS archive (\url{https://gnss.nps.gov/doi-gnss/Rinex/}) and CHAIN (\url{https://www.chain-project.net/}).

Dual-frequency GNSS observations from the northern and southern polar networks were used for the regional TEC/ROTI analyses. These data are available through EarthScope (\url{https://data.unavco.org/archive/gnss/}), CDDIS (\url{https://cddis.nasa.gov/archive/gnss/}), BKG (\url{https://igs.bkg.bund.de/root_ftp/}) and Geoscience Australia (\url{https://data.gnss.ga.gov.au/}).

High-rate ground GNSS observations and the amplitude and phase scintillation indices $S_4$ and $\sigma_\phi$ are available from CHAIN (\url{https://www.chain-project.net/data/gps/ismr/}) and INGV eSWua (\url{https://www.eswua.ingv.it/}).

The LEO-RO SNR and excess-phase products used here are available through UCAR CDAAC's NOAA near-real-time collections for PlanetiQ (\url{https://data.cosmic.ucar.edu/gnss-ro/planetiq/noaa/nrt/}) and Spire (\url{https://data.cosmic.ucar.edu/gnss-ro/spire/noaa/nrt/}), the CDAAC post-processed collection for TerraSAR-X (\url{https://data.cosmic.ucar.edu/gnss-ro/tsx/postProc/level1b/}), and EUMETSAT for MetOp (\url{https://user.eumetsat.int/catalogue/EO:EUM:DAT:METOP:GRASL1}). The corresponding PlanetiQ and Spire raw-data collections are archived by NOAA/NCEI at \url{https://doi.org/10.25921/7xdk-vy89} and \url{https://doi.org/10.25921/m2te-hp45}, respectively.

SuperDARN radar observations are available through \url{https://superdarn.ca/data-access/}, and PFISR electron-density measurements through the SRI Madrigal database (\url{https://data.amisr.com/madrigal/}). The IRI model and IGRF-14 coefficients are available from the IRI project (\url{https://irimodel.org/}) and NOAA/NCEI (\url{https://www.ncei.noaa.gov/products/international-geomagnetic-reference-field}), respectively.

Space-weather indices are available from OMNIWeb (\url{https://omniweb.gsfc.nasa.gov/}), GFZ (\url{https://kp.gfz.de/en/}), WDC Kyoto (\url{https://wdc.kugi.kyoto-u.ac.jp/}) and the Polar Cap index service (\url{https://pcindex.org/}). Map data are from Natural Earth (\url{https://www.naturalearthdata.com/downloads/}).

The March 2025 YunYao electron-density profiles were supplied on request by Tianjin Yunyao Aerospace Technology Co., Ltd. and were not part of its public release at the time of analysis. Requests for the original profiles should be directed to the provider through its data platform (\url{https://open.tjyyspace.com/#/dataDownload}).
\endgroup

% Keep the availability heading with its paragraph.
\par
\ifdim\dimexpr\pagegoal-\pagetotal\relax<5\baselineskip
  \newpage
\fi

\section*{Code availability}

{The analysis and plotting code is available at \url{https://doi.org/10.5281/zenodo.21675357}.} LuNART is available at \url{https://github.com/NavsasPolito/LuNART}.
\endgroup

\section*{Author contributions}
R.Y. coordinated the study, led the analysis and figure preparation, and drafted the manuscript. Z.L. and X.M. contributed to the processing and analysis of ground-based GNSS and LEO-RO data. Y.M. and W.G. contributed to LOS geometry calculations for the LuGRE observations. A.N., A.M. and F.D. contributed expertise on the LuGRE measurements and data processing. C.C., L.S. and L.A. contributed to the ionospheric background and physical interpretation. L.-T.H. participated in discussions of the manuscript concept and structure. X.Z. provided project funding and led discussions of the study's scientific contribution. All authors contributed to the discussion and revision of the manuscript.

\begingroup
\def\blpage#1{#1}
\bibliography{LuGRE_R5}% 
\endgroup

\clearpage
\setcounter{figure}{0}
\renewcommand{\figurename}{Extended Data Fig.}
\renewcommand{\theHfigure}{ExtendedData.\arabic{figure}}

% Full-page display for the approved tall ED1; other pages retain their original layout.
\newgeometry{left=25mm,right=25mm,top=15mm,bottom=20mm,bindingoffset=0mm}
\begin{figure}[p]
\centering
\includegraphics[width=\textwidth]{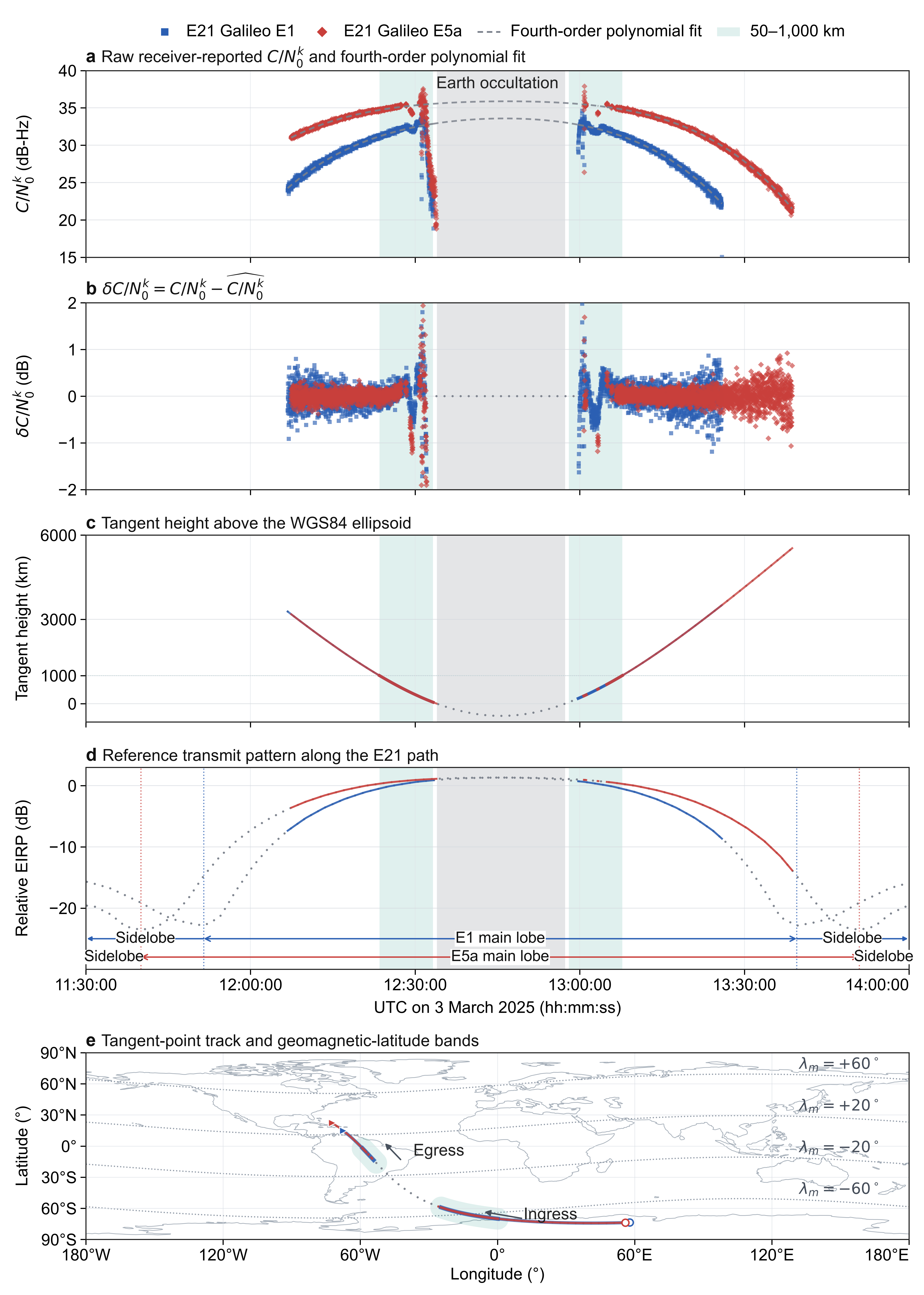}
% R5: inherited revision-link colour removed; preserve paragraph spacing.
\caption{{\textbf{Galileo E21 E1/E5a as an example of arc reconstruction and fourth-order polynomial detrending.}
Measured segments before and after Earth occultation on 3 March 2025 are combined into one complete arc, with a single fourth-order polynomial fitted independently to each signal (\textbf{a}). The slow background can include range-dependent loss and transmitter-gain variations illustrated by the Galileo reference pattern (\textbf{d}) \cite{Menzione2024GalileoAntenna}. Subtracting the fit reveals rapid $\delta C/N_0$ fluctuations in the 50--1{,}000~km analysis intervals near high reference gain (\textbf{b}, shown with a $\pm2$~dB zoom). Tangent height and track are calculated from receiver and satellite positions (\textbf{c,e}). Smaller residuals above 1{,}000~km are assessed as background in Extended Data Fig.~\ref{edfig:cn0-quality-background}.}}
\label{edfig:cn0-detrending-op38-e21}
\end{figure}
\clearpage
\restoregeometry

% Approved QC/background figure: separate page at the same readable width as ED1.
\newgeometry{left=25mm,right=25mm,top=15mm,bottom=20mm,bindingoffset=0mm}
\begin{figure}[p]
\centering
\includegraphics[width=\textwidth]{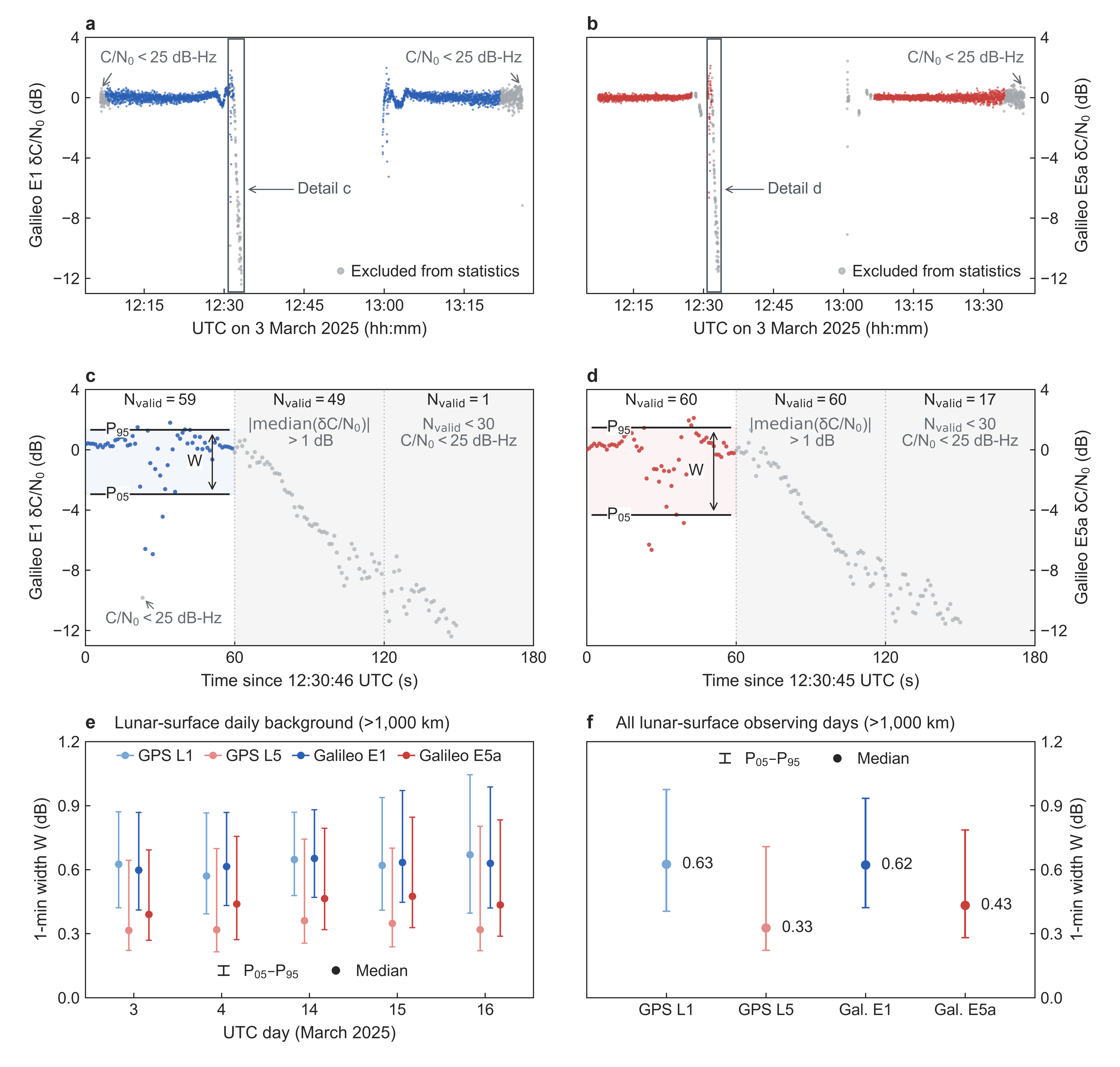}
\caption{{\textbf{Galileo E21 E1/E5a quality-screening example and estimation of background noise fluctuations.}
Measurements with $C/N_0<25$~dB-Hz are excluded first to limit fluctuations associated with unstable weak-signal tracking. The remaining $\delta C/N_0$ observations are grouped into 60-s windows. Retained windows must satisfy $|\mathrm{median}(\delta C/N_0)|\leq1$~dB to limit residual offsets from incomplete background removal, and contain at least 30 valid 1-Hz observations ($N_{\rm valid}\geq30$) to avoid estimates based on sparse sampling (\textbf{\mbox{a--d}}; grey points are excluded). Windows failing either condition are excluded in full. These checks apply to all $\delta C/N_0$ statistics. For valid windows, the central 90\% width, $W=P_{95}(\delta C/N_0)-P_{05}(\delta C/N_0)$, characterizes fluctuation strength while reducing the influence of extreme tails. Applying the same screening above 1{,}000~km tangent height across the nearly 93~h lunar-surface campaign gives \revSJTU{larger background $C/N_0$ deviations at L1/E1 than at L5/E5a} (\textbf{e,f}). \revSJTU{These background $C/N_0$ fluctuations typically span approximately 0.5~dB and probably reflect noise-induced deviations in the onboard $C/N_0$ estimates.} The E1 and E5a ionospheric fluctuations in \textbf{c,d} have widths $W>4$~dB, well above those caused by background noise.}}
\label{edfig:cn0-quality-background}
\end{figure}
\clearpage
\restoregeometry

\begin{figure}[p]
\centering
\includegraphics[width=\textwidth]{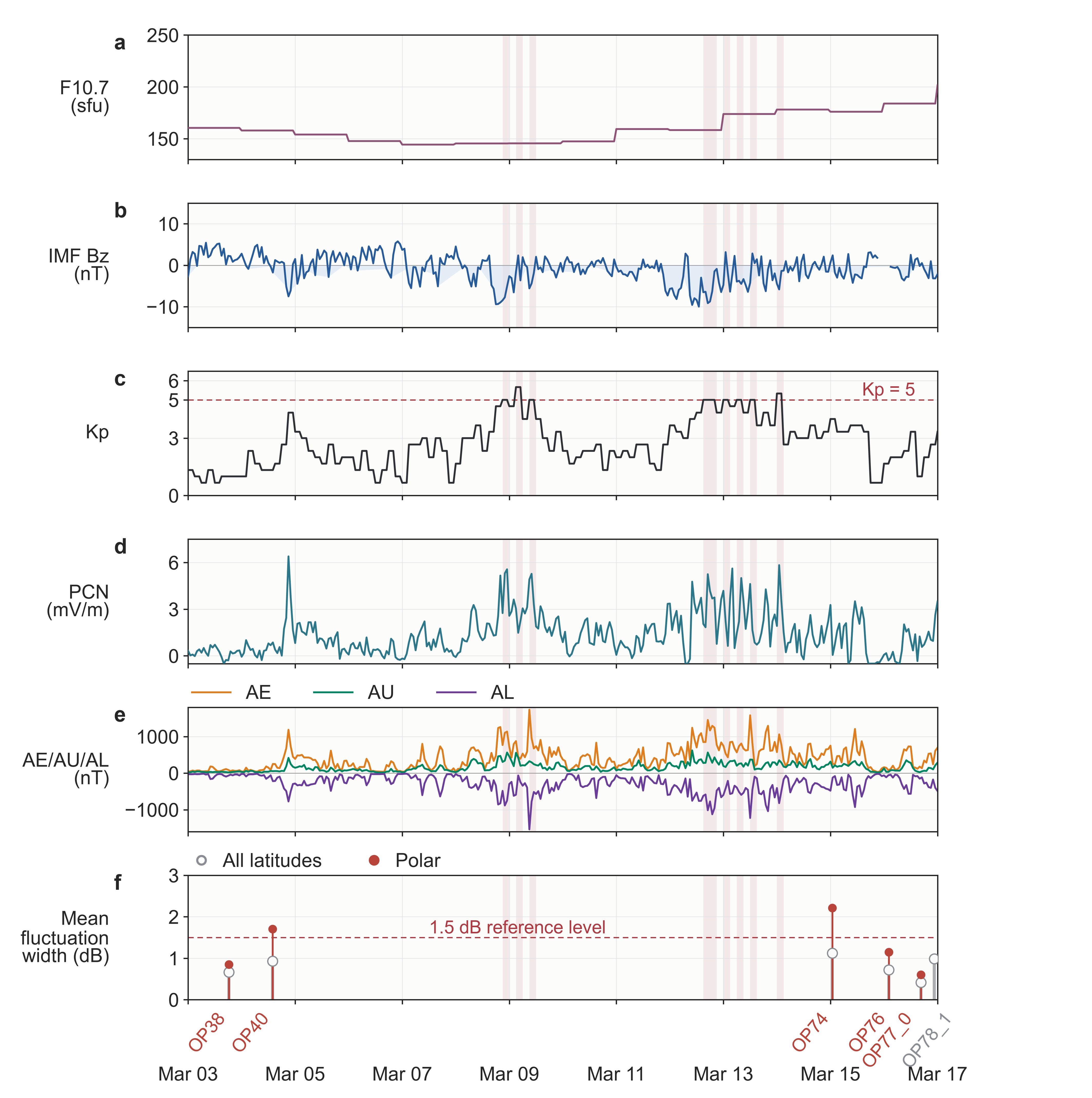}
\caption{{
\textbf{Polar amplitude enhancement during relatively quiet geomagnetic conditions.}
\textbf{a--e}, F10.7 solar radio flux, interplanetary magnetic-field $B_z$, \revSJTU{the planetary K-index (Kp)}, \revSJTU{the polar cap activity index for Northern Hemisphere (PCN)} and auroral-electrojet indices (AE, AU and AL) \revSJTU{\cite{Kauristie2017GeomagneticIndices}}, respectively, during 3--17 March 2025. Pale red shading marks Kp~$\geq5$; the dashed line in \textbf{c} marks Kp~=~5.
\textbf{f}, Mean width of $\delta C/N_0$ fluctuation for each lunar surface operation. Filled red symbols show polar observations and open grey symbols show all latitudes; the dashed line marks the 1.5-dB reference level.
Polar-cap and auroral activity increased intermittently during the campaign. LuGRE observed pronounced polar amplitude fluctuations during the relatively quiet geomagnetic intervals sampled.
}}
\label{edfig:solarwind-lugre-context}
\end{figure}

\begin{figure}[p]
\centering
\includegraphics[width=\textwidth]{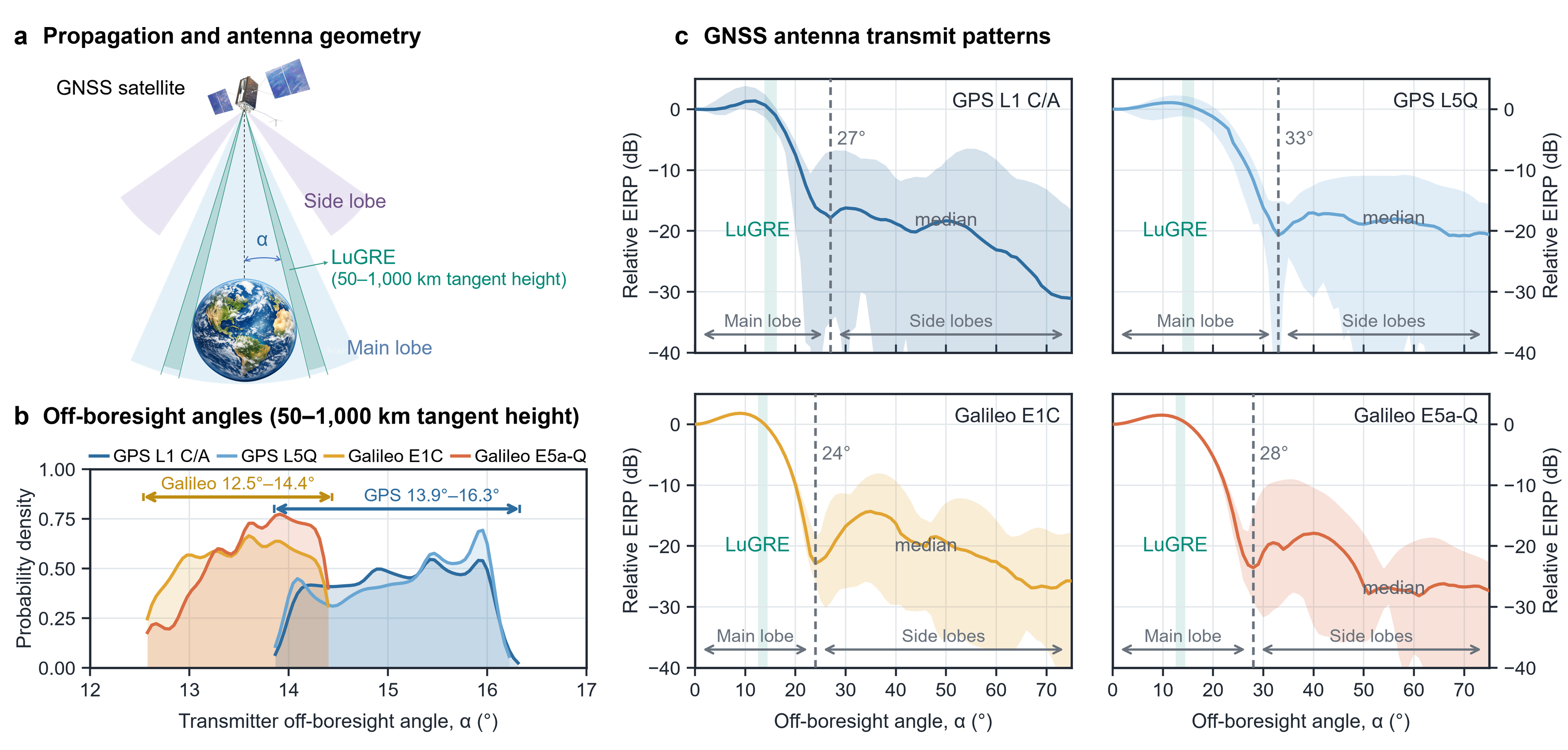}
% R5: inherited revision-link colour removed; preserve paragraph spacing.
\caption{{
\textbf{LuGRE ionospheric observations sample the high-gain portions of GNSS main lobes.}
\textbf{a,b}, Earth-limb geometry and transmitter off-boresight angles $\alpha$, measured from the Earth-centre direction, for LuGRE observations at tangent heights of 50--1{,}000~km. The sampled angles span approximately $13.9^\circ$--$16.3^\circ$ for GPS and $12.5^\circ$--$14.4^\circ$ for Galileo.
\textbf{c}, In the illustrated reference transmit patterns, the main lobes extend to approximately $27^\circ$ for GPS L1, $33^\circ$ for GPS L5, $24^\circ$ for Galileo E1 and $28^\circ$ for Galileo E5a, with sidelobes at larger $\alpha$ \cite{Marquis2016GPSBlockIIR,Montenbruck2024GPSAntenna,Menzione2024GalileoAntenna}. The sampled LuGRE directions lie well within the relatively high-gain main-lobe regions.
}}
\label{edfig:main-lobe-antenna}
\end{figure}

\begin{figure}[p]
\centering
\includegraphics[width=\textwidth]{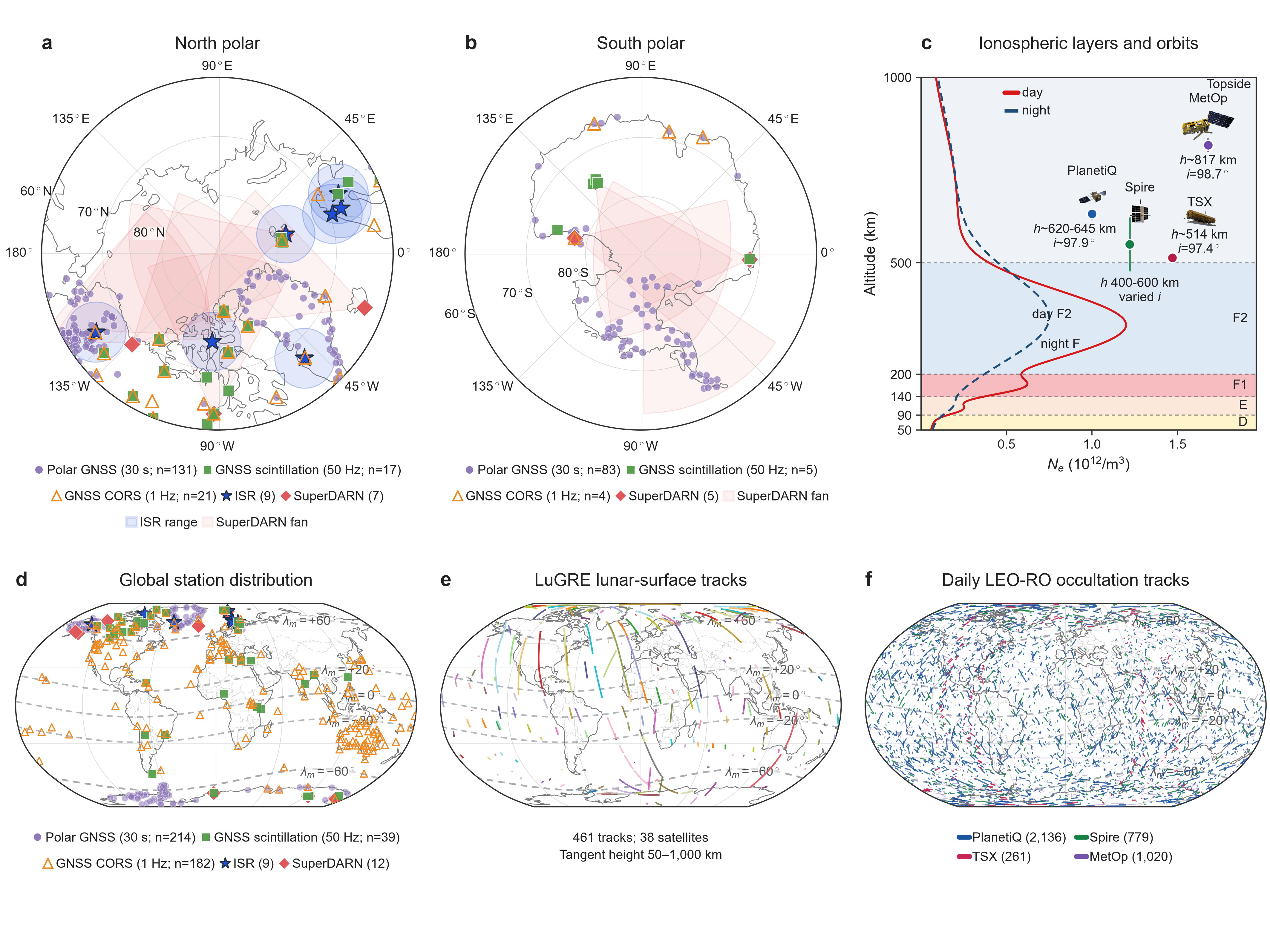}
\caption{{
\textbf{Observing networks and sampling coverage for ionospheric comparisons.}
\textbf{a,b}, Northern and southern polar networks. Purple circles mark polar GNSS stations with 30-s observations; green squares mark 50-Hz GNSS scintillation stations; orange triangles mark 1-Hz GNSS continuously operating reference stations (CORS) contributing to the amplitude comparison. Blue stars and red diamonds mark incoherent-scatter radars (ISR) and coherent-radars (SuperDARN), respectively. Shaded circles and fans show schematic radar coverage.
\textbf{c}, Schematic daytime and nighttime electron-density profiles, ionospheric layers and representative LEO mission altitudes. Coloured points and vertical bars indicate reference orbital heights and schematic height ranges; $h$ and $i$ denote altitude and inclination.
\textbf{d}, Global distribution of the observing networks. Legend numbers give total counts within each map.
\textbf{e}, LuGRE lunar-surface tangent-point tracks on 3--4 and 14--16 March 2025, at tangent heights of 50--1{,}000~km. The 461 satellite--signal tracks involve 38 GNSS satellites, distinguished by colour.
\textbf{f}, Selected LEO-RO daily tracks during 14--15 March 2025: PlanetiQ (blue), Spire (green), TerraSAR-X (TSX; magenta) and MetOp (purple). Numbers in the legend give track counts.
Dashed lines in \textbf{d--f} mark geomagnetic latitudes of $0^\circ$, $\pm20^\circ$ and $\pm60^\circ$.
}}
\label{edfig:observing-context}
\end{figure}

\begin{figure}[p]
\centering
\includegraphics[width=\textwidth]{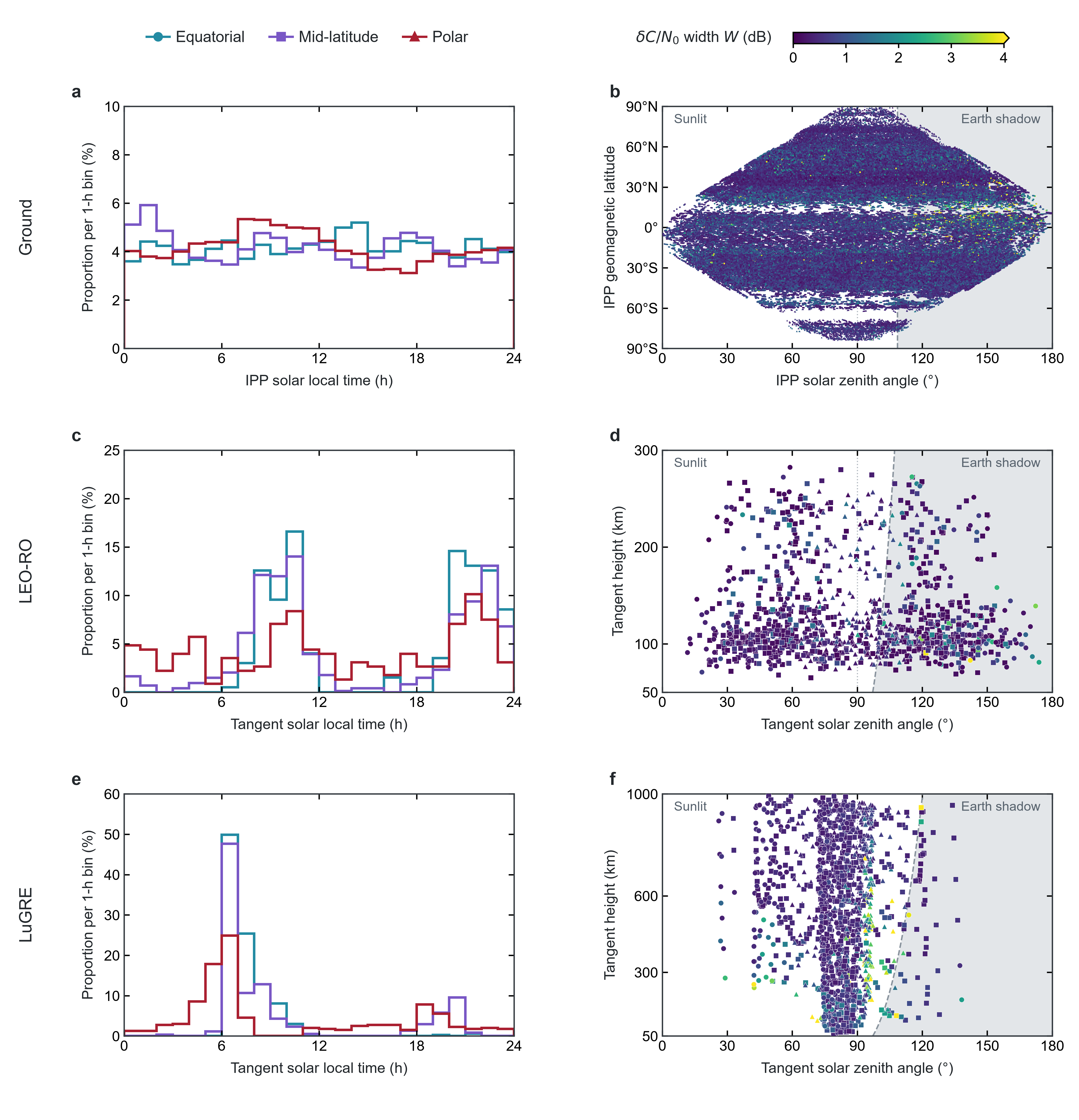}
\caption{{
\textbf{Local-time and illumination coverage of ground, LEO and lunar observations.}
Ground sampling spans local time relatively evenly in all three latitude bands (\textbf{a}). LEO-RO observations are concentrated in the 06--12 and 18--24 local-time sectors, especially at equatorial and mid-latitudes (\textbf{c}). Lunar sampling has a morning peak in all three bands: equatorial and mid-latitude observations cluster around 06--10 local time, while polar coverage is broader (\textbf{e}). Equatorial plasma bubbles (EPBs) and their associated irregularities commonly develop after \revSJTU{local sunset at the magnetic equator and ionospheric altitudes} \cite{Spogli2023EPB}. Sparse lunar coverage of this interval may undersample these irregularities, contributing to the weaker observed equatorial amplitude fluctuations.
Illumination is evaluated at the 350-km Ground IPP or the actual LEO-RO/LuGRE tangent point (\textbf{b,d,f}), with each point representing a 1-min window coloured by its width $W$. Pronounced lunar polar fluctuations occur under sunlit conditions; LEO-RO also shows large fluctuations in Earth's shadow, and Ground samples both illumination conditions.
}}
\label{edfig:local-time-illumination}
\end{figure}

\begin{figure}[tbp]
\centering
\includegraphics[width=\textwidth]{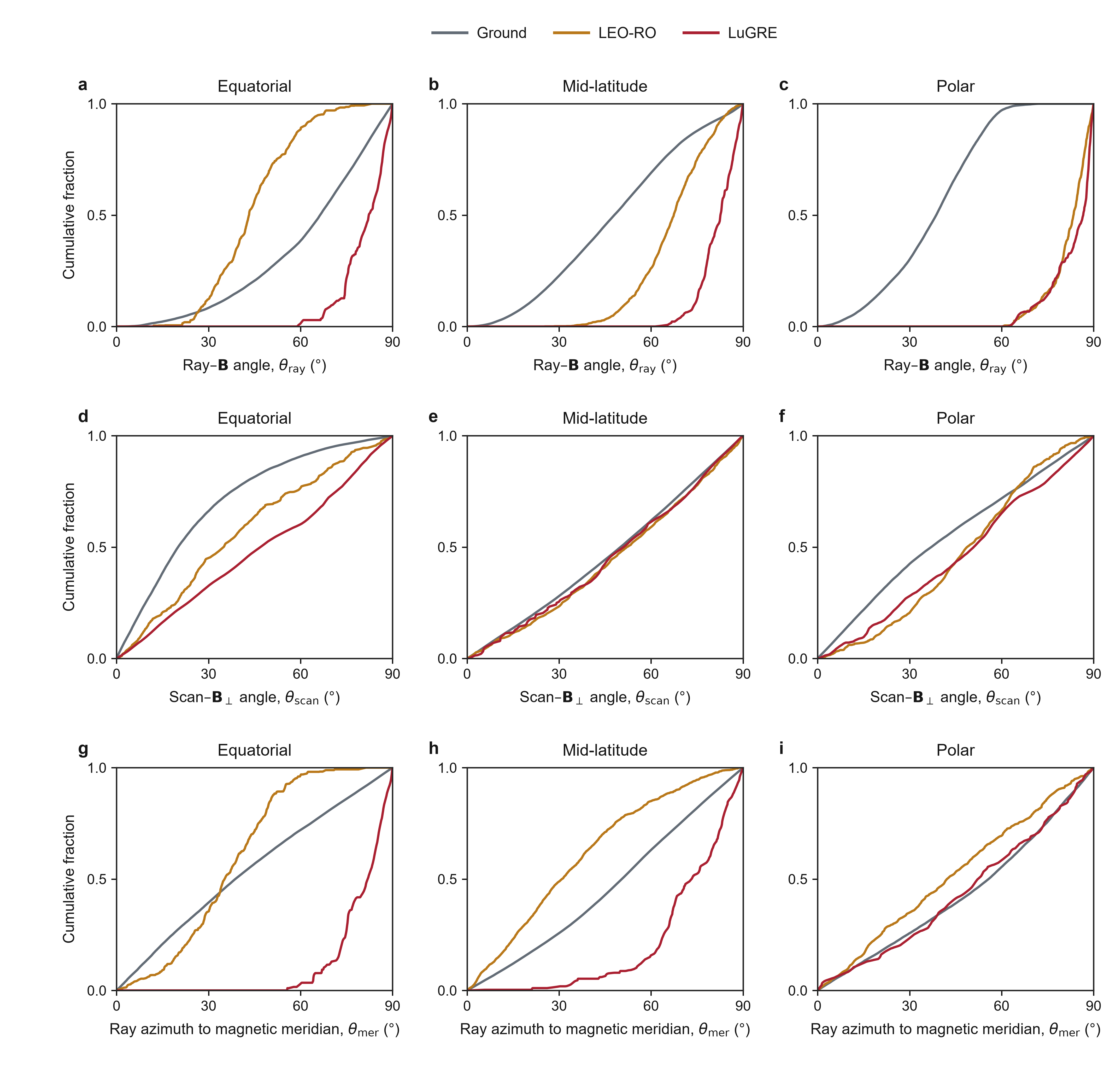}
\caption{{
\textbf{Lunar ray orientations differ most from near-Earth geometries at lower latitudes.}
Angles are calculated relative to the local IGRF-14 magnetic field at the Ground IPP or LEO-RO/LuGRE tangent point: ray--field angle $\theta_{\rm ray}$ (\textbf{a--c}), scan--projected-field angle $\theta_{\rm scan}$ (\textbf{d--f}) and magnetic-meridian azimuth $\theta_{\rm mer}$ (\textbf{g--i}).
LuGRE rays are nearly perpendicular to the field in all three latitude bands. Polar LEO-RO rays share this geometry, whereas equatorial and mid-latitude LuGRE rays have larger ray--field and magnetic-meridian angles than near-Earth observations. Scan-angle distributions are similar among platforms at mid-latitudes and overlap substantially between LEO-RO and LuGRE in the polar band; platform differences are larger at equatorial latitudes.
The stronger polar lunar scan-angle association in Fig.~\ref{fig4} thus occurs within angular coverage that substantially overlaps LEO-RO.
}}
\label{edfig:geometry-angles}
\end{figure}

\begin{figure}[tbp]
\centering
\includegraphics[width=\textwidth]{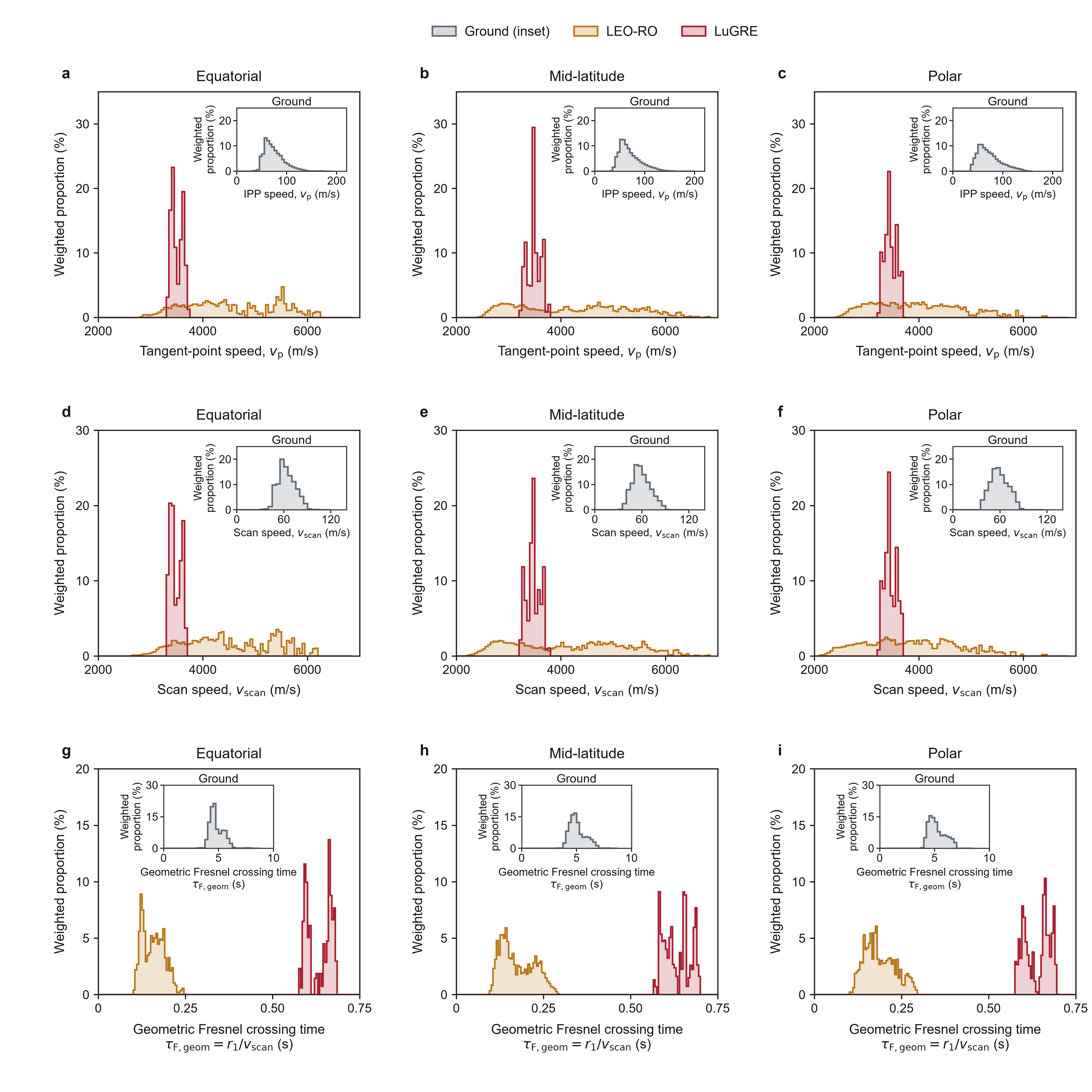}
\caption{{
\textbf{Ground, LEO and lunar observations sample complementary motion and propagation scales.}
Reference-point speeds (\textbf{a--c}) and transverse scan speeds (\textbf{d--f}) are narrowly concentrated around 3{,}500~m/s for LuGRE, span a broader range for LEO-RO and are much smaller for Ground.
Despite near-space scan speeds of a similar order, the larger lunar Fresnel scale gives longer geometric crossing times than LEO-RO; Ground has the longest times (\textbf{g--i}). These estimates describe geometrical motion without irregularity drift, which would change the effective relative sampling. Together with their different Fresnel scales and angular responses, the platforms provide complementary temporal views of ionospheric structure.
}}
\label{edfig:scan-speed-fresnel}
\end{figure}

\begin{figure}[tbp]
\centering
\includegraphics[width=\textwidth]{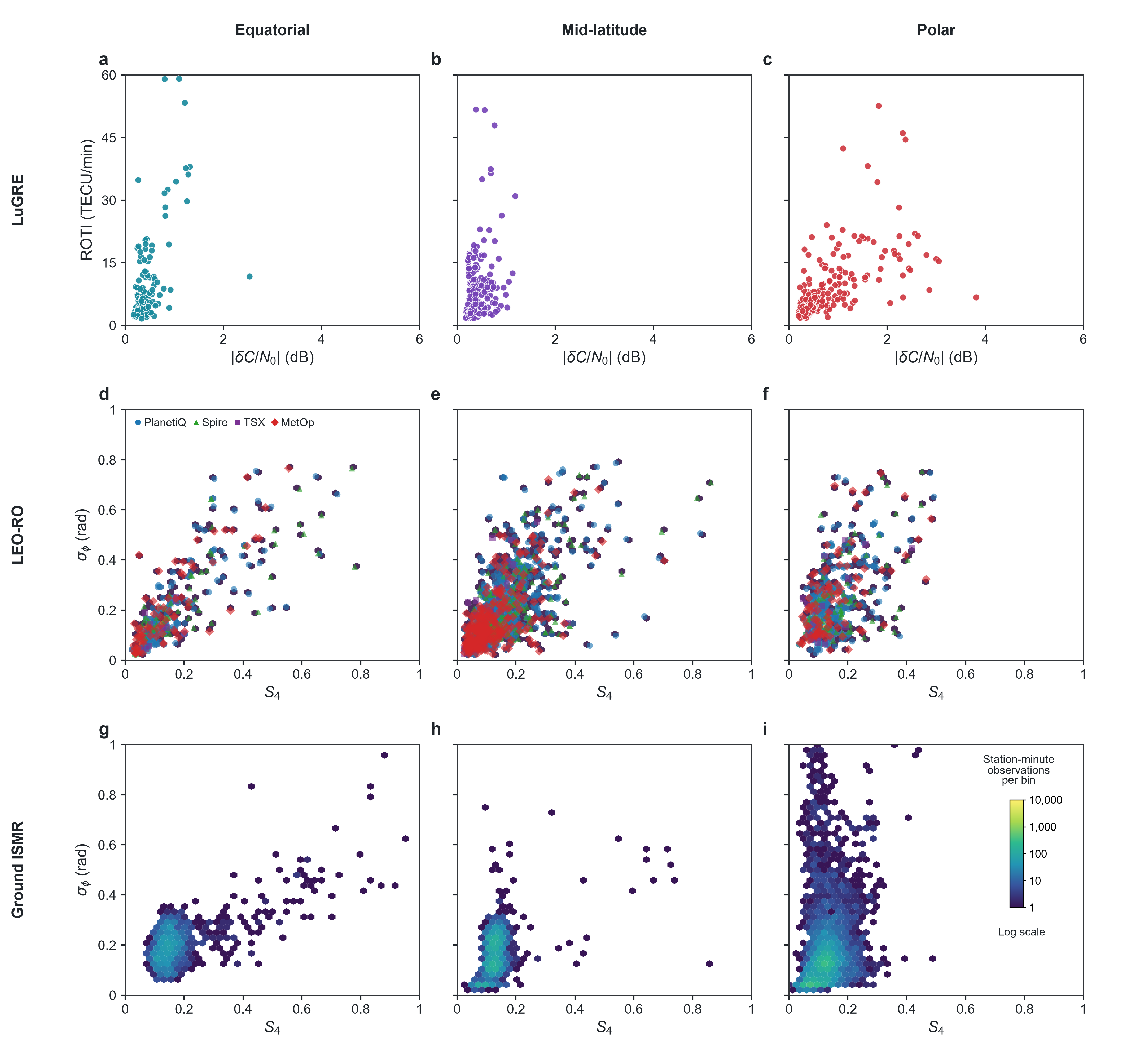}
\caption{{
\textbf{Polar lunar amplitude enhancement alongside near-Earth amplitude and phase responses.}
LuGRE's 1-Hz $C/N_0$ fluctuations and phase-derived ROTI characterize amplitude and electron-content variability (\textbf{a--c}). They provide a reference comparison with the conventional high-rate scintillation indices $S_4$ and $\sigma_\phi$ from LEO-RO (\textbf{d--f}) \cite{Chang2025PlanetiQSpireRO} and Ground (\textbf{g--i}), with distinct physical meanings and temporal sampling.
LuGRE shows broader polar amplitude fluctuations, whereas large ROTI also occurs at lower latitudes. LEO-RO amplitude and phase indices tend to increase together across the latitude bands. Ground polar observations often combine weak amplitude scintillation with substantial phase variability.
}}
\label{edfig:ROTI_S4}
\end{figure}

\end{document}